\documentclass[journal]{IEEEtran}
\usepackage[cmex10]{amsmath}
\usepackage{amssymb}
\usepackage{mathtools}
\usepackage{array}

\usepackage{algpseudocode,algorithm,algorithmicx}
\usepackage{float}
\usepackage{subfigure}
\usepackage{tabularx}
\usepackage{mathrsfs} 
\usepackage{psfrag}
\usepackage{tikz}
\usetikzlibrary{chains, arrows,shapes,positioning,arrows,decorations.markings,calc, spy}
\usepackage{pgfplots}
\pgfplotsset{compat=newest}
\usepgfplotslibrary{groupplots}
\usepackage[pdfusetitle,hidelinks, pdfauthor={Kilian Roth}, pdfsubject={A Comparison of Hybrid Beamforming and Digital Beamforming with Low-Resolution ADCs for Multiple Users and Imperfect CSI}]{hyperref}
\usepackage{gensymb}
\usepackage{xcolor}
\usepackage[shortcuts, acronym]{glossaries}

\pgfplotsset{
    mark repeat*/.style={
        scatter,
        scatter src=x,
        scatter/@pre marker code/.code={
            \pgfmathtruncatemacro\usemark{
                or(mod(\coordindex,#1)==0, (\coordindex==(\numcoords-1))
            }
            \ifnum\usemark=0
                \pgfplotsset{mark=none}
            \fi
        },
        scatter/@post marker code/.code={}
    }
}

\definecolor{GTOBMIX1}{rgb}{0,0.80,0.20}
\definecolor{GTOBMIX2}{rgb}{0,0.60,0.40}
\definecolor{GTOBMIX3}{rgb}{0,0.40,0.60}
\definecolor{GTOBMIX4}{rgb}{0,0.20,0.80}

\definecolor{BTORMIX1}{rgb}{0.125, 0,0.875}
\definecolor{BTORMIX2}{rgb}{0.25, 0,0.75}
\definecolor{BTORMIX3}{rgb}{0.375,0,0.625}
\definecolor{BTORMIX4}{rgb}{0.50,0,0.50}
\definecolor{BTORMIX5}{rgb}{0.625,0,0.375}
\definecolor{BTORMIX6}{rgb}{0.75,0,0.25}
\definecolor{BTORMIX7}{rgb}{0.875,0,0.125}

\definecolor{BTORMIX16}{rgb}{0.143, 0,0.857}
\definecolor{BTORMIX26}{rgb}{0.286, 0,0.714}
\definecolor{BTORMIX36}{rgb}{0.429,0,0.571}
\definecolor{BTORMIX46}{rgb}{0.571,0,0.429}
\definecolor{BTORMIX56}{rgb}{0.714,0,0.286}
\definecolor{BTORMIX66}{rgb}{0.857,0,0.143}

\definecolor{BTORMIX12}{rgb}{0.333, 0,0.667}
\definecolor{BTORMIX22}{rgb}{0.667, 0,0.333}

\definecolor{BTOGMIX12}{rgb}{0, 0.333,0.667}
\definecolor{BTOGMIX22}{rgb}{0, 0.667,0.333}

\definecolor{BTOGMIX1}{rgb}{0, 0.125, 0.875}
\definecolor{BTOGMIX2}{rgb}{0, 0.25, 0.75}
\definecolor{BTOGMIX3}{rgb}{0, 0.375, 0.625}
\definecolor{BTOGMIX4}{rgb}{0, 0.50, 0.50}
\definecolor{BTOGMIX5}{rgb}{0, 0.625, 0.375}
\definecolor{BTOGMIX6}{rgb}{0, 0.75 ,0.25}
\definecolor{BTOGMIX7}{rgb}{0, 0.875, 0.125}

\definecolor{BTOGMIX16}{rgb}{0, 0.143,0.857}
\definecolor{BTOGMIX26}{rgb}{0, 0.286,0.714}
\definecolor{BTOGMIX36}{rgb}{0, 0.429,0.571}
\definecolor{BTOGMIX46}{rgb}{0, 0.571,0.429}
\definecolor{BTOGMIX56}{rgb}{0, 0.714,0.286}
\definecolor{BTOGMIX66}{rgb}{0, 0.857,0.143}

\begin{document}
%
\title{A Comparison of Hybrid Beamforming and Digital Beamforming with Low-Resolution ADCs for Multiple Users and Imperfect CSI}
%
%
%
%
\author{Kilian~Roth,~\IEEEmembership{Member,~IEEE,}
		Hessam~Pirzadeh,~\IEEEmembership{Member,~IEEE,}
		A. Lee~Swindlehurst~\IEEEmembership{Fellow,~IEEE,}
        Josef~A.~Nossek,~\IEEEmembership{Life Fellow,~IEEE}%
\thanks{K. Roth is with Next Generation and Standards, Intel Deutschland GmbH, Neubiberg 85579, Germany (email: kilian.roth@intel.com)}%
\thanks{K. Roth and J. A. Nossek are with the Department of Electrical and Computer Engineering, Technical University Munich, Munich 80290, Germany (email: kilian.roth@tum.de; josef.a.nossek@tum.de)}%
\thanks{J. A. Nossek is also with Department of Teleinformatics Engineering, Federal University of Ceara, 60020-180 Fortaleza, Brazil}%
\thanks{H. Pirzadeh and A. L. Swindlehurst are with the Center for Pervasive Communications and Computing, University of California, Irvine, CA 92697 USA. (e-mail: hpirzade@uci.edu; swindle@uci.edu).}%
\thanks{A. L. Swindlehurst is also with Institute for Advanced Study, Technical University of Munich, Munchen 80333, Germany}}%

\maketitle

\newacronym{A/D}{A/D}{Analog/Digital}
\newacronym[plural=ADCs,firstplural=Analog-to-Digital-Converters (ADCs)]{ADC}{ADC}{Analog-to-Digital-Converter}
\newacronym{AGC}{AGC}{Automatic Gain Control}
\newacronym{AQNM}{AQNM}{Additive Quantization Noise Model}
\newacronym{BB}{BB}{BaseBand}
\newacronym{CIR}{CIR}{Channel Impulse Response}
\newacronym{CMOS}{CMOS}{Complementary Metal–Oxide–Semiconductor}
\newacronym{DBF}{DBF}{Digital BeamForming}
\newacronym{DMRS}{DMRS}{DeModulation Reference Signals}
\newacronym{EVM}{EVM}{Error Vector Magnitude}
\newacronym{FDM}{FDM}{Frequency Domain Multiplex}
\newacronym{HBF}{HBF}{Hybrid BeamForming}
\newacronym{ISM}{ISM}{Industrial, Scientific and Medical}
\newacronym{LA}{LA}{Limiting Amplifier}
\newacronym[plural=LOs,firstplural=Local Oscillators (LOs)]{LO}{LO}{Local Oscillators}
\newacronym{LTE}{LTE}{Long Term Evolution}
\newacronym{MIMO}{MIMO}{Multiple Input Multiple Output}
\newacronym{MINLP}{MINLP}{Mixed Integer Non-Linear Programing}
\newacronym{mmWave}{mmWave}{millimeter Wave}
\newacronym{MMSE}{MMSE}{Minimum Mean Square Error}
\newacronym{MSE}{MSE}{Mean Square Error}
\newacronym{MU-MIMO}{MU-MIMO}{Multi User - Multiple Input Multiple Output}
\newacronym{NLP}{NLP}{NonLinear Programing}
\newacronym{NR}{NR}{New Radio}
\newacronym{OFDM}{OFDM}{Orthogonal Frequency Domain Multiplexing}
\newacronym{PDP}{PDP}{Power Delay Profile}
\newacronym[plural=PAs,firstplural=Power Amplifiers (PAs)]{PA}{PA}{Power Amplifier}
\newacronym{Q}{Q}{Quantization}
\newacronym{RF}{RF}{Radio Frequency}
\newacronym{RFE}{RFE}{Radio Front-End}
\newacronym{Rx}{Rx}{receiver}
\newacronym[plural=SISOs,firstplural=Single Input Single Output (SISOs)]{SISO}{SISO}{Single Input Single Output}
\newacronym{SNR}{SNR}{Signal to Noise Ratio}
\newacronym[plural=SCs,firstplural=Sub-Carriers (SCs)]{SC}{SC}{Sub-Carrier}
\newacronym{Tx}{Tx}{transmitter}
\newacronym[plural=UE,firstplural=User Equipment (UE)]{UE}{UE}{User Equipment}
\newacronym{ULA}{ULA}{Uniform Linear Array}
\newacronym{VGA}{VGA}{Variable Gain Amplifier}

\begin{abstract}
For 5G it will be important to leverage the available millimeter wave spectrum.
To achieve an approximately omnidirectional coverage with a similar effective antenna aperture compared to state of the art cellular systems, an antenna array is required at both the mobile and basestation. Due to the large bandwidth and inefficient amplifiers available in CMOS for mmWave, the analog front-end of the receiver with a large number of antennas becomes especially power hungry.
Two main solutions exist to reduce the power consumption: hybrid beam forming and digital beam forming with low resolution Analog to Digital Converters (ADCs).
In this work we compare the spectral and energy efficiency of both systems under practical system constraints.
We consider the effects of channel estimation, transmitter impairments and multiple simultaneous users.
Our power consumption model considers components reported in literature at 60 GHz. 
In contrast to many other works we also consider the correlation of the quantization error, and generalize the modeling of it to non-uniform quantizers and different quantizers at each antenna. 
The result shows that as the SNR gets larger the ADC resolution achieving the optimal energy efficiency gets also larger. 
The energy efficiency peaks for 5 bit resolution at high SNR, since due to other limiting factors the achievable rate almost saturates at this resolution. 
We also show that in the multi-user scenario digital beamforming is in any case more energy efficient than hybrid beamforming. 
In addition we show that if different ADC resolutions are used we can achieve any desired trade-offs between power consumption and rate close to those achieved with only one ADC resolution. 
\end{abstract}
%
\IEEEpeerreviewmaketitle
%
%
%
%
\begin{figure*}[!t]
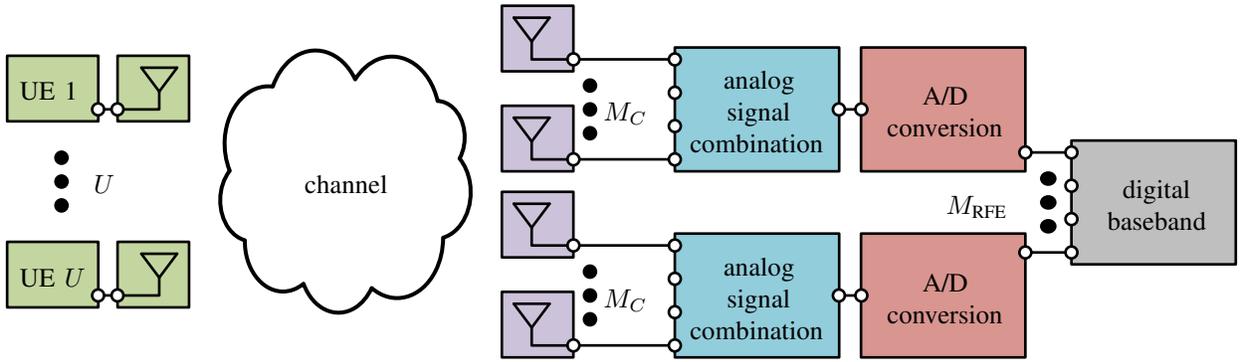

\begin{center}
\normalsize
	\ifCLASSOPTIONdraftcls
		\input{./Introduction/pics/OverviewModelpsfrag_1col.tex}
	\else
		\input{./Introduction/pics/OverviewModelpsfrag_2col.tex}
	\fi
	\caption{System Model with $U$ UEs with 1 antenna and $M_{C}$ antennas at each of the $M_{\text{RFE}}$ RF chains at the basestations. Number of receive Antennas $M_R$ is equal to $M_{C}\times M_{\text{RFE}}$}
	\label{fig:SystemModel}
\hrulefill
\end{center}
\vspace*{-0.5cm}
\end{figure*}
\section{Introduction}
The use of the available bandwidth in the frequency range of 6 to 100 GHz is considered to be an essential part of the next generation mobile broadband standard 5G \cite{FIVEDISRUPTIVE}. 
Due to the propagation condition at these frequencies, this technology is especially attractive for high data rate, shorter range wireless communication. 
This frequency range is referred to as \ac*{mmWave}, even though it contains the lower 
centimeter wave range. In recent years, the spectrum and the availability of consumer grade systems at mmWave frequencies has led to a huge increase in academic and industrial research. 
However, to fully leverage the spectrum while being power-efficient, the \ac*{BB} and \ac*{RFE} capabilities must be drastically changed from current state of the art cellular devices. 

The use of high carrier frequencies above 6 GHz will go hand in hand with the implementation of large antenna arrays 
\cite{FIVEDISRUPTIVE}, \cite{WHATWILL5GBE}. 
The support of a large number of antennas at the mobile and base station requires a new \ac*{RFE} design.
To attain a similar link budget, the effective antenna aperture of a \ac*{mmWave} system must be comparable to current
systems operating at carrier frequencies below 6 GHz. Therefore, an antenna array at 
both the base and mobile station is unavoidable. Since the antenna gain and thus the directivity 
increases with the aperture, an antenna array is the only solution to achieve a high effective aperture while maintaining an omnidirectional coverage. 

\subsection{Related Work}
Current \ac*{LTE} systems have a limited amount of antennas at the base and mobile stations. Since the bandwidth is relatively narrow,
the power consumption of a receiver \ac*{RF} chain with a high resolution \ac*{ADC} at each antenna is still feasible.
For future \ac*{mmWave} mobile broadband systems, a much larger bandwidth \cite{NGNM5GWHITE} and a much large number of antennas
are being considered \cite{FIVEDISRUPTIVE}. 
The survey in \cite{ADCSURVEY} shows that \acp*{ADC}
with a high sampling frequency and a standard number of effective bits of resolution (6-10) consume a considerable amount of power.
Consequently, the power consumption of the \ac*{ADC} can be considered as the bottleneck of the receiver \cite{1BITMIMOINFO1}. 

The use of a large antenna array combined with a large bandwidth is a huge challenge for the hardware implementation; essentially
the power consumption will limit the design space. At the moment, analog/hybrid beamforming is considered 
as a possible solution to reduce the power consumption.
Analog or hybrid beamforming systems strongly depend on the calibration of the analog components. Another major disadvantage is the 
large overhead associated with the alignment of the Tx and Rx beams of the base and mobile station. 
Specifically, if high gain is needed, the beamwidth is small and thus the acquisition and constant alignment of the optimal beams in a 
dynamic environment is very challenging \cite{CellSearchDirectionalmmW, rappaport2014millimeter, 1BITCAPHEATH}.

The idea of hybrid beamforming is based on the concept of phased array antennas commonly used in radar applications \cite{MAILLOUXPHASEDARRAY}. 
Due to the reduced power consumption, it is also seen as a possible solution for \ac*{mmWave} mobile broadband communication\cite{HBFMAG}. 
If the phased array approach is combined with digital beamforming, the phased array approach might also be feasible for non-static or quasi-static scenarios.
In \cite{Kong:EECS-2014-191}, it was shown that considering the inefficiency of \ac*{mmWave} amplifiers and the high insertion loss of \ac*{RF} phase shifters, it
is better to perform the phase shifting in the baseband. The power consumption associated with both cases is comparable, as long as the 
number of antennas per RF-chain remains relatively small. 

Another option to reduce the power consumption while keeping the number of antennas constant is to reduce the power consumption of the \acp*{ADC} by
reducing their resolution. This can also be combined with hybrid beamforming. Some of these evaluations consider only the extreme case of 1-bit quantization 
\cite{1BITCAPHEATH, 1BITIDEA2, 1BITMIMOINFO1, YLI2017}. In \cite{AQNMAMINE, QINGOPAQNM} the \ac*{A/D} conversion is modeled as a linear
stochastic process. Low resolution \ac*{A/D} conversion combined with \ac*{OFDM} in an uplink scenario are considered in \cite{ADCOFDM, OFDM1BIT}.

In \cite{HLOWADC, ACHCDCCOMP} hybrid beamforming with low resolution \ac*{A/D} conversion was considered.
The energy efficiency / spectral efficiency trade-off of fully-connected hybrid and digital beamforming with low resolution \acp*{ADC} is assessed in 
\cite{ACHCDCCOMP}. But in contrast as shown in the system diagram in Fig. \ref{fig:SystemModel}, we consider
a hybrid beamforming system that has exclusive antennas per \ac*{RF}-chain (aka. sub-array hybrid beamforming).
In this work we concentrated on effects of the hardware constraints at the receiver, thus we assumed the transmitter to be ideal.
In \cite{ACHCDCCOMP}, a fully-connected hybrid beamforming system is used, which has a large
additional overhead
 associated with an increased number of phase shifters and larger power combiners. Also in this case additional amplifiers to compensate for the insertion-loss of the \ac*{RF} phase shifters and combiners are required.
In \cite{ONERLOWRES}, analog beamforming is compared with digital beamforming in terms of power efficiency.

The authors of \cite{YOO2006, LEE2012} both analyzed the effect of imperfect channel knowledge on the achievable rate. The channel estimation error is treated as additional noise added to the system. 
We will use a similar model to include the channel estimation error into our analysis. Since we we have a system involving multiple user with different receive power, we treat the effect of each users separately. 

\subsection{Contribution}
The contribution of this work can be summarized in the following bulletpoints:
\begin{itemize}
	\item Achievable rate analysis for digital and hybrid beamforming systems with low resolution \acp*{ADC} in a multi-user, multipath scenario. In addition the effects of transmitter impairments, channel estimation errors and having a mixed \ac*{ADC} resolutions are considered.
	\item Analyzing the channel estimation error considering the reference signal patterns already agreed upon for 3GPP NR (aka. 5G).
	\item Showing the energy efficiency - spectral efficiency trade-off considering the power consumption of the receiver RF front-end. 
	\item Generalizing the \ac*{AQNM} to include the effects for quantization error correlation, non-uniform quantization and different \acp*{ADC} at each antenna.
\end{itemize}

\subsection{Notation}
Throughout the paper we use boldface lower and upper case letters to represent column vectors and matrices.
The term $a_{m,l}$ is the element on row $m$ and column $l$ of matrix $\boldsymbol{A}$ and $a_m$ is the $m$th element of vector $\boldsymbol{a}$. 
The expressions $\boldsymbol{A}^*$, $\boldsymbol{A}^T$, $\boldsymbol{A}^H$, and $\boldsymbol{A}^{-1}$ 
represent the complex conjugate, the transpose, the Hermitian, and the inverse of the matrix $\boldsymbol{A}$.
The symbol $\boldsymbol{R}_{\boldsymbol{a}\boldsymbol{b}}$ is the correlation matrix of vector $\boldsymbol{a}$ and $\boldsymbol{b}$ defined as
$\mathbb{E}[\boldsymbol{a}\boldsymbol{b}^H]$.
The Discrete Fourier Transformation (DFT) $\mathcal{F}(\cdot)$ and its inverse $\mathcal{F}^{-1}(\cdot)$ and the Fourier transformation $\mathscr{F}\{\cdot\}$ and its inverse
$\mathscr{F}^{-1}\{\cdot\}$ are also used.

\section{Signal Model}
The system model in Fig. \ref{fig:SystemModel} gives
a general overview of both investigated systems. For $M_C = 1$
the block analog signal combination is just connecting the
input to the output. For $M_C > 1$ this block contains an analog
phase shifter for each signal followed by a power combiner.

The symbols $\boldsymbol{x}_u[n]$, $\boldsymbol{\eta}_u[n]$,
$\boldsymbol{H}_u[l]$, $\boldsymbol{\eta}_R[n]$, and $\boldsymbol{y}[n]$ represent the complex valued transmit
signal of user $u$, the imperfections of the transmitter of user $u$, channel from user $u$ to the basestation, the noise at the receiver, and the receive signal of the system, respectively. We assume that there are $U$ users with $M_T$ antennas each and a basestation with $M_R$ receive antennas.
The receive signal $\boldsymbol{y}[n]$ is defined as
\begin{equation}
	\boldsymbol{y}[n] = \sum\limits_{u=1}^{U} \sqrt{P_u} \sum\limits_{l=0}^{L_u} \boldsymbol{H}_u[l] ( \boldsymbol{x}_u[n-l] + \boldsymbol{\eta}_u[n-l]) + \boldsymbol{\eta}_R[n],
\end{equation}
where $P_u$ is the transmit power of user $u$ and $L_u$ is the length of the channel in samples from user $u$ to the basestation. 
The transmitter impairments $\boldsymbol{\eta}_u[n]$ are modeled as circular symmetric complex Gaussian noise with zero mean and covariance equal to $\sigma_{\text{EVM}}^2$. Including the transmit power
$P_u$, this is the classical \ac*{EVM} definition only considering transmitter impairments \cite{reynaert2006rf}.

Since all noise contributions are Gaussian we can combine them to form a combined noise $\boldsymbol{\eta}^\prime_R[n]$ equal to
\begin{equation}
	\boldsymbol{\eta}^\prime_R[n] =  \sum\limits_{u=1}^{U} \sqrt{P_u} \sum\limits_{l=0}^{L_u} \boldsymbol{H}_u[l] \boldsymbol{\eta}_u[n-l] + \boldsymbol{\eta}_R[n].
\end{equation}
The receive signal is then reduced to 
\begin{equation}
	\boldsymbol{y}[n] = \sum\limits_{u=1}^{U} \sqrt{P_u} \sum\limits_{l=0}^{L_u} \boldsymbol{H}_u[l] \boldsymbol{x}_u[n-l]+ \boldsymbol{\eta}^\prime_R[n].
\end{equation}

We restrict the system to have $M_{C}$ antennas
exclusively connected to one RF front-end chain (see Fig. \ref{fig:SystemModel}). Therefore, the matrix modeling the analog combining at the receiver $\boldsymbol{W}_R$ has the form
\begin{equation}
	\boldsymbol{W}_R = 
	\begin{bmatrix}
		\boldsymbol{w}_R^1  & \boldsymbol{0}_{M_C}   & \hdots  & \boldsymbol{0}_{M_C}  \\
		\boldsymbol{0}_{M_C} & \boldsymbol{w}_R^2  & \ddots  & \boldsymbol{0}_{M_C}  \\
		\vdots & \ddots  & \ddots &\vdots    \\
		\boldsymbol{0}_{M_C} & \hdots  & \boldsymbol{0}_{M_C} & \boldsymbol{w}_R^{M_{RFE}} \\
	\end{bmatrix} \in \mathbb{C}^{M_R\times M_{RFE}},
\end{equation}
where the vector $\boldsymbol{w}_R^i$ is the analog beamforming vector of the $i$th RF chain. 
We also restrict our evaluation to the case where each RF chain is connected to the same number of antennas $M_{C}$. 
The vectors $\boldsymbol{w}_R^i$ and $\boldsymbol{0}_{M_C}$ have dimension $M_{C}$. 
The receiver signal ${y}_C[n]$ after the analog combining $\boldsymbol{y}_C[n]$ is then
\begin{equation}
	\boldsymbol{y}_C[n] =  \boldsymbol{W}^H_R \boldsymbol{y}[n].
\end{equation}
For the case of digital beamforming the matrix $\boldsymbol{W}_R$ is simply replace by an identity matrix with the same dimensions. 

For the case of \ac*{DBF}, we study cases where the \acp*{ADC} have either uniform resolution or a mixture of different resolutions. 
In our evaluation, we will restrict our attention to the following type of scenarios: $M_h$ \acp*{ADC} with a higher resolution $b_h$ and $M_l$ \acp*{ADC} with a lower resolution $b_l$. 
The channel model assumes the same average receive power at each antenna for each user. 
This means that the high resolution \acp*{ADC} can be allocated to any $M_h$ antennas, and the remaining antennas to the \acp*{ADC} with lower resolution. 
In practical scenarios it would be very difficult to adaptively allocate different \acp*{ADC} to different \ac*{RF} chains, since it takes a non-negligible amount of time to perform the switching.
Furthermore, we do not expect the received power to be different on average for different antennas, so allocating the $M_h$ high resolution \acp*{ADC} to an arbitrary subset of the antennas is a reasonable approach.

\subsection{Channel Model}
The measurements in \cite{mmMAGIC_D2_1} show that for channels at 60 GHz, an exponential \ac*{PDP} sufficiently approximates a real world scenario
\begin{equation}
	\boldsymbol{H}[l] = \frac{1}{\sqrt{M_T} } \alpha(l) \boldsymbol{a}_r(\phi_r(l)) \boldsymbol{a}_t^T(\phi_t(l)).
\end{equation}
The phase shift between
the signal at adjacent antenna elements at the receiver and transmitter $\phi_r(l)$ and $\phi_t(l)$ of path $l$ depend on the angle of arrival $\theta_r(l)$ and departure $\theta_t(l)$
\begin{equation}
	\boldsymbol{a}_r^T(\phi_r(l)) = \left[1,e^{j \phi_r(l)},e^{j2 \phi_r(l)},\cdots,e^{j(M_r - 1) \phi_r(l)}\right].
\end{equation}
Here we assume, that at delay $l$ only one ray arrives at the receiver. The complex gain of the ray $\alpha(l)$ is assumed to be circular symmetric Gaussian distributed with zero mean and a variance defined according to
\begin{equation}
	v_l = \mathbb{E}\left[\left\vert\alpha(l)\right\vert^2\right]  = e^{-\beta l}.
\end{equation}
The parameter $\beta$ defines how fast the power decays in relation to the delay.
The other parameters of the model are the maximum channel length in samples $L$ and the number of present channel taps $P$. 
This means for any channel realization, only $P$ elements of the $L \times 1$ vector of variances $\boldsymbol{v}$ are non-zero.
We will normalize the variance vector as follows:
\begin{equation}
	\boldsymbol{v}_n = \frac{\boldsymbol{v}}{\vert\vert\boldsymbol{v}\vert\vert^2}.
\end{equation}

The \ac*{SNR} $\gamma_u$ per user $u$ is defined as as
\begin{equation}
	\gamma_u = \frac{P_u~\mathbb{E} \left[\left\vert \left\vert \sum\limits^{L_u}_{l=0} \boldsymbol{H}_u[l] \boldsymbol{x}_u[n-l] \right\vert \right\vert^2_2 \right]}{ \mathbb{E} [\left\vert \left\vert \boldsymbol{\eta}_R[n] \right\vert \right\vert^2_2 ]}.
	\label{eq:SNRDef}
\end{equation}
This formula describes the average \ac*{SNR} at each antenna. It is important to note that the expectation takes the realization of the channel and realizations of $\boldsymbol{x}_i[n]$ into account.

\subsection{Analytic MSE of frequency domain channel estimation with time-frequency interpolation}
Assuming perfect synchronization of the timing and carrier frequency, the \ac*{OFDM} receive signal $Y_{(k, \ell, m)}$ of subcarrier $k$, \ac*{OFDM} symbol $\ell$ and antenna $m$ can be written as
\begin{equation}
	Y_{(k, \ell, m)} = H_{(k, \ell, m)} X_{(k, \ell, m)} + \eta_{(k, \ell, m)},
\end{equation}
where we assume that the \ac*{CIR} is shorter than the cyclic prefix, and $H_{(k, \ell, m)}$, $X_{(k, \ell, m)}$ and $\eta_{(k, \ell, m)}$ are the channel,
 transmit signal and white Gaussian noise of the system, respectively.
To include channel estimation errors into the rate analysis, we evaluate the theoretical channel estimation performance. Since frequency domain channel estimation is equivalent transform domain channel estimation in \ac*{OFDM}, we
reformulate the theoretical \ac*{MSE} expressions for our system. In \cite{Biagini2014} the \ac*{MSE} for the reference signal pattern of \ac*{LTE} is calculated. Time-frequency filters are used to interpolate the channel 
estimate between the position of the reference symbols. The theoretical \ac*{MSE} is identical with the version calculated based on channel realizations. A 2-D time-frequency interpolation method based on a \ac*{MMSE} criteria as
described in \cite{CEMMSE} is identified as the solution with the best performance.

In contrast, we use a 3-D time-frequency-space filter for smoothing of the estimate in the frequency domain. It is important to note that this technique assumes knowledge of the following statistical channel parameters:
\begin{itemize}
	\item Doppler shift
	\item Delay spread
	\item Signal power of each user
	\item Noise power
	\item Spatial correlation
\end{itemize}
Since in addition we consider a \ac*{MU-MIMO} scenario we need to ensure that different users have orthogonal reference sequences. In particular, we will assume that the training sequences are orthogonal.
We assume that orthogonality is ensured by \ac*{FDM} and a cyclic shift of the reference symbols. 
Therefore, the following calculation is done for each user, and thus no user index is included to simplify the notation.

Assuming a reference symbol is present on subcarrier $q$ and symbol time $p$ we multiply the signal with the known reference signal to obtain the corresponding channel estimate for antenna $m$
\begin{equation}
	\hat{H}_{(p, q, m)} = Y_{(p, q, m)}  X^*_{(p, q)} = H_{(p, q, m)} + \eta_{(p, q, m)},
\end{equation}
where we assume that $\left\vert X^*_{(p, q)}\right\vert = 1$. 
By combining the channel estimates for all resource elements on $K$ subcarriers, $L$ symbols and $M$ antennas we get
\begin{equation}
	\hat{\boldsymbol{h}}_r = \left[ \hat{H}_{(1, 1, 1)}, \hat{H}_{(2, 1, 1)}, \cdots, \hat{H}_{(K-1, L, M)},  \hat{H}_{(K, L, M)} \right]^T.
\end{equation}
For all positions where no reference signals were sent the corresponding element of $\hat{\boldsymbol{h}}_r$ is set equal to zero.
The set $\mathbb{P}$ contains the indices of the reference symbols in $\hat{\boldsymbol{h}}_r$.

Applying the matrices for interpolation and smoothing in time $\boldsymbol{A}_t$, frequency $\boldsymbol{A}_f$ and space $\boldsymbol{A}_s$ 
we get the overall estimate of the channel at each position 
\begin{equation}
	\hat{\boldsymbol{h}} = \boldsymbol{A}_{stf} \hat{\boldsymbol{h}}_r = \left(\boldsymbol{A}_s \otimes \boldsymbol{A}_t \otimes \boldsymbol{A}_f\right) \hat{\boldsymbol{h}}_r.
\end{equation}
We choose these interpolation matrices separately for each dimension to reduce the complexity.
In general to achieve the theoretical optimal performance these interpolation matrices have to be chosen according
to the covariance matrix of the channel, which might not be separable. As shown in \cite{CEMMSE} for the time-frequency case this leads to a minimal performance loss, but with significantly lower complexity. 
In many cases the covariance is unknown, and one would need to generate the interpolation martrices based on some model for the covariance, whose parameters would also then have to be estimated.

The \ac*{MSE} of the estimate $\hat{\boldsymbol{h}}$ compared to the actual channel $\boldsymbol{h}$ can be calculate as
\begin{equation}
\begin{gathered}
	\frac{1}{KLM}\mathbb{E}\left[\left\vert\left\vert\hat{\boldsymbol{h}} - \boldsymbol{h}\right\vert\right\vert^2\right] = \\
	\frac{1}{KLM} \left(\mathbb{E}\left[\hat{\boldsymbol{h}}^H \hat{\boldsymbol{h}}\right] - 2 \Re\left(\mathbb{E}\left[\hat{\boldsymbol{h}}^H \boldsymbol{h}\right]\right) + \mathbb{E}\left[ \boldsymbol{h}^H \boldsymbol{h}\right]\right).
\label{eq:MSECH1}
\end{gathered}
\end{equation}
We split the term in \eqref{eq:MSECH1} into three components and calculated them separately.

The third component can be calculated as
\begin{equation}
	\mathbb{E}\left[ \boldsymbol{h}^H \boldsymbol{h}\right]  = 
	\text{tr}(\boldsymbol{R}_{\boldsymbol{h}\boldsymbol{h}})  = \text{tr}( \boldsymbol{R}^s_{\boldsymbol{h}\boldsymbol{h}} \otimes \boldsymbol{R}^t_{\boldsymbol{h}\boldsymbol{h}} \otimes \boldsymbol{R}^f_{\boldsymbol{h}\boldsymbol{h}} ).
	\label{eq:Term3MSECH}
\end{equation}
The covariance matrices $\boldsymbol{R}^t_{\boldsymbol{h}\boldsymbol{h}}$,  $\boldsymbol{R}^f_{\boldsymbol{h}\boldsymbol{h}}$ and $\boldsymbol{R}^s_{\boldsymbol{h}\boldsymbol{h}}$ are the time, frequency and spatial covariance matrices of the channel. It is important to keep in mind that this separation might not be possible across all domains, dependent on the channel statistics. The channel model chosen in this work allows this separation. 

The first component can be calculated as
\begin{equation}
	\begin{gathered}
	\mathbb{E}\left[\hat{\boldsymbol{h}}^H \hat{\boldsymbol{h}}\right] 
	= \text{tr}\left( \boldsymbol{A}_{stf} \mathbb{E}\left[\hat{\boldsymbol{h}}_r \hat{\boldsymbol{h}}_r^H\right] \boldsymbol{A}_{stf}^H\right), \\
	\mathbb{E}\left[\hat{\boldsymbol{h}}_r \hat{\boldsymbol{h}}_r^H\right] = \sum\limits_{p1 \in \mathbb{P}} \sum\limits_{p2 \in \mathbb{P}}
	\left[\boldsymbol{R}_{\boldsymbol{h}\boldsymbol{h}} + \boldsymbol{R}_{\boldsymbol{\eta}\boldsymbol{\eta}}\right]_{p1,p2} \boldsymbol{e}_{p1} \boldsymbol{e}^T_{p2},
	\label{eq:Term1MSECH}
	\end{gathered}
\end{equation}
where $\boldsymbol{R}_{\boldsymbol{\eta}\boldsymbol{\eta}}$ is the covariance matrix of the noise across space, time and frequency.
The vector $\boldsymbol{e}_p$ is a vector with only zeros, and a one at the $p$th position.
We assume it can be also be separated into the submatrices for space, time and frequency in the same was as the channel:
\begin{equation}
	\boldsymbol{R}_{\boldsymbol{\eta}\boldsymbol{\eta}} = \boldsymbol{R}^s_{\boldsymbol{\eta}\boldsymbol{\eta}} \otimes \boldsymbol{R}^t_{\boldsymbol{\eta}\boldsymbol{\eta}} \otimes \boldsymbol{R}^f_{\boldsymbol{\eta}\boldsymbol{\eta}}.
\end{equation}
 
The second component of \eqref{eq:MSECH1} can be calculated in a similar fashion as the previous one
\begin{equation}
	\begin{gathered}
	\mathbb{E}\left[\hat{\boldsymbol{h}}_r^H \boldsymbol{h} \right] = \text{tr}\left( \mathbb{E}\left[\boldsymbol{h} \hat{\boldsymbol{h}}_r^H \right] \boldsymbol{A}_{tf}^H\right), \\
	\mathbb{E}\left[\boldsymbol{h} \hat{\boldsymbol{h}}_r^H\right] =
	\sum\limits_{p \in \mathbb{P}} \boldsymbol{R}_{\boldsymbol{h}\boldsymbol{h}} \boldsymbol{e}_p \boldsymbol{e}^T_p,
	\label{eq:Term2MSECH}
	\end{gathered}
\end{equation}
using that fact that the noise has zero mean.

 Plugging \eqref{eq:Term3MSECH}, \eqref{eq:Term1MSECH} and \eqref{eq:Term2MSECH} into \eqref{eq:MSECH1} we get the analytic \ac*{MSE} as
\begin{equation}
\begin{gathered}
	\frac{1}{KLM}\mathbb{E}\left[\left\vert\left\vert\hat{\boldsymbol{h}} - \boldsymbol{h}\right\vert\right\vert^2\right] = \frac{1}{KLM}\Bigg[\\
	\text{tr}\left( \boldsymbol{A}_{stf} \left( \sum\limits_{p1 \in \mathbb{P}} \sum\limits_{p2 \in \mathbb{P}}
	\left[\boldsymbol{R}_{\boldsymbol{h}\boldsymbol{h}} + \boldsymbol{R}_{\boldsymbol{\eta}\boldsymbol{\eta}}\right]_{p1,p2} \boldsymbol{e}_{p1} \boldsymbol{e}^T_{p2} \right)
 	\boldsymbol{A}_{stf}^H\right) \\
	-~2\Re\left(\text{tr}\left( \left(\sum\limits_{p \in \mathbb{P}} \boldsymbol{R}_{\boldsymbol{h}\boldsymbol{h}} \boldsymbol{e}_p
	\boldsymbol{e}^T_p \right)\boldsymbol{A}_{stf}^H\right)\right) \\
	+~\text{tr}\left(\boldsymbol{R}_{\boldsymbol{h}\boldsymbol{h}}\right)\Bigg].
\end{gathered}
\end{equation}
If we can decompose the matrices $\boldsymbol{A}_{stf}$, $\boldsymbol{R}_{\boldsymbol{h}\boldsymbol{h}}$ and $\boldsymbol{R}_{\boldsymbol{\eta}\boldsymbol{\eta}}$ into the Kronecker product of three matrices
the computation of the MSE can be simplified to:
\begin{equation}
	\frac{1}{KLM}\mathbb{E}\left[\left\vert\left\vert\hat{\boldsymbol{h}} - \boldsymbol{h}\right\vert\right\vert^2\right] =
	\frac{1}{KLM}\left[C1 - 2 \Re(C2) + C3\right],
\end{equation}
with the components $C1$, $C2$ and $C3$ defined as:
\begin{equation}
\begin{gathered}
	C1 = \sum\limits_{p1 \in \mathbb{P}} \sum\limits_{p2 \in \mathbb{P}} \bigg([\boldsymbol{R}^s_{\boldsymbol{h}\boldsymbol{h}}]_{m1, m2}[\boldsymbol{R}^t_{\boldsymbol{h}\boldsymbol{h}}]_{\ell1, \ell2}[\boldsymbol{R}^f_{\boldsymbol{h}\boldsymbol{h}}]_{k1, k2}  +\\
	[\boldsymbol{R}^s_{\boldsymbol{\eta}\boldsymbol{\eta}}]_{m1, m2}[\boldsymbol{R}^t_{\boldsymbol{\eta}\boldsymbol{\eta}}]_{\ell1, \ell2}[\boldsymbol{R}^f_{\boldsymbol{\eta}\boldsymbol{\eta}}]_{k1, k2} \bigg) \\
	\left([\boldsymbol{A}_s]^H_{m2} [\boldsymbol{A}_s]_{m1} \right)\left([\boldsymbol{A}_t]^H_{\ell2} [\boldsymbol{A}_t]_{\ell1}\right)\left([\boldsymbol{A}_f]^H_{k2} [\boldsymbol{A}_f]_{k1} \right)\\
	C2 = \sum\limits_{p \in \mathbb{P}} [\boldsymbol{A}_s^H\boldsymbol{R}^s_{\boldsymbol{h}\boldsymbol{h}}]_{m,m}[\boldsymbol{A}_t^H\boldsymbol{R}^t_{\boldsymbol{h}\boldsymbol{h}}]_{\ell,\ell}[\boldsymbol{A}_f^H\boldsymbol{R}^f_{\boldsymbol{h}\boldsymbol{h}}]_{k,k}\\
	C3 = \text{tr}\left( \boldsymbol{R}^s_{\boldsymbol{h}\boldsymbol{h}}\right)\text{tr}\left( \boldsymbol{R}^t_{\boldsymbol{h}\boldsymbol{h}}\right)\text{tr}\left( \boldsymbol{R}^f_{\boldsymbol{h}\boldsymbol{h}}\right),
\end{gathered}
\end{equation}
where $m$, $m1$, $m2$, $k$, $k1$, $k2$, $\ell$, $\ell 1$ and $\ell 2$ are the space, frequency and time indices corresponding the position of the reference symbols. 

\begin{figure}
	\centering
	\input{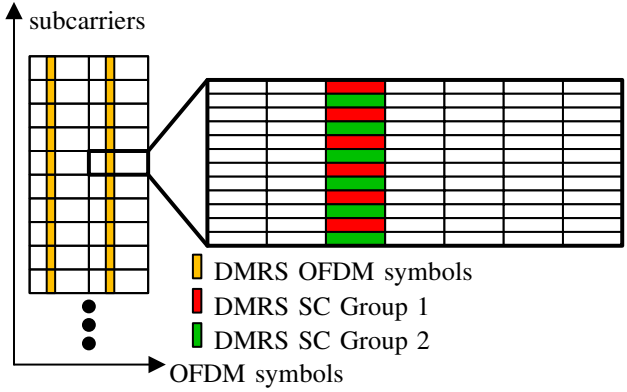}
	\caption{3GPP NR OFDM type one reference signal pattern for up to 4 UEs.}
	\label{fig:RSPattern}
\end{figure}

The interpolation/spatial smoothing matrices $\boldsymbol{A}_t$ and $\boldsymbol{A}_f$ are chosen according to 
\cite{CEMMSE} based on knowledge of the \ac*{SNR}, the delay spread including a model for the \ac*{PDP} and the Doppler spread. 
Since all these parameters are estimated and afterwards generated according to a model, they will never
exactly match the actual \ac*{PDP} and Doppler spread. This introduces a model mismatch that is included in our evaluation. 

The time and frequency covariance matrices $\boldsymbol{R}_t$ and $\boldsymbol{R}_f$ can be calculated according to the actual \ac*{PDP} 
and the Doppler shift including the corresponding model as shown in \cite{Biagini2014}. Based on the correlation matrix $\boldsymbol{R}_{\text{CIR}}$ of the \ac*{CIR} we can
calculate the correlation matrix in the frequency domain $\boldsymbol{R}_f$ as
\begin{equation}
	\boldsymbol{R}_f = \boldsymbol{W} \boldsymbol{R}_{\text{CIR}} \boldsymbol{W}^H,
\end{equation}
where $\boldsymbol{W}$ is the matrix corresponding to a DFT transformation.

\ifCLASSOPTIONdraftcls
\begin{figure}
\begin{center}
\begin{tikzpicture}
    \begin{axis}[width=8.8cm, height=6cm, xlabel={SNR [dB]},  legend cell align=left, axis on top, ymode = log, grid, 
                  		ylabel={Channel Estimation MSE}, xmin=-30, xmax=30, ymax=10^(2.5), ymin=10^(-3.5), ytick={-30, -20, -10, 0, 10, 20, 30},
                  		ytick={100, 10, 1, 0.1, 0.01, 0.001}]
                  		\addplot [no markers, very thick, red] table [x = x, y=y] {./SignalModel/PlotData/SNR_vs_MSE.txt};
                  		
     \end{axis}
    \begin{axis}[width=8.8cm, height=6cm, xlabel={SNR [dB]},  legend cell align=left, axis on top, grid, axis y line*=right, axis x line=none,
                  		ylabel={SNR degredation}, legend style={at={(0.85,0.97)}}, legend cell align=left, legend columns=1, xmin=-30, xmax=30, ymin = 0.25, ymax = 3.25,
                  		ytick={3, 2.5, 2, 1.5, 1, 0.5}]
                  		\addlegendimage{no markers, very thick, red}\addlegendentry{MSE}
                  		\addplot [no markers, very thick, blue, dashed] table [x = x, y=y] {./SignalModel/PlotData/SNR_vs_SNR_degrad.txt};
                  		\addlegendentry{SNR degradation}
                  		
     \end{axis}
\end{tikzpicture}
\vspace*{-0.5cm}
\end{center}
\caption{Channel estimation MSE and resulting SNR degradation dependent on input SNR.}
\label{fig:CHMSE}
\end{figure}
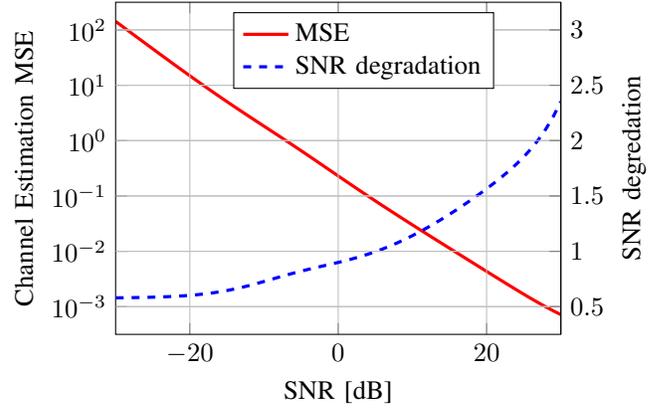
\else
\begin{figure}
\begin{center}
\begin{tikzpicture}
    \begin{axis}[width=7.5cm, height=6cm, xlabel={SNR [dB]},  legend cell align=left, axis on top, ymode = log, grid, 
                  		ylabel={Channel Estimation MSE}, xmin=-30, xmax=30, ymax=10^(2.5), ymin=10^(-3.5), ytick={-30, -20, -10, 0, 10, 20, 30},
                  		ytick={100, 10, 1, 0.1, 0.01, 0.001}]
                  		\addplot [no markers, very thick, red] table [x = x, y=y] {./SignalModel/PlotData/SNR_vs_MSE.txt};
                  		
     \end{axis}
    \begin{axis}[width=7.5cm, height=6cm, xlabel={SNR [dB]},  legend cell align=left, axis on top, grid, axis y line*=right, axis x line=none,
                  		ylabel={SNR degredation}, legend style={at={(0.85,0.97)}}, legend cell align=left, legend columns=1, xmin=-30, xmax=30, ymin = 0.25, ymax = 3.25,
                  		ytick={3, 2.5, 2, 1.5, 1, 0.5}]
                  		\addlegendimage{no markers, very thick, red}\addlegendentry{MSE}
                  		\addplot [no markers, very thick, blue, dashed] table [x = x, y=y] {./SignalModel/PlotData/SNR_vs_SNR_degrad.txt};
                  		\addlegendentry{SNR degradation}
                  		
     \end{axis}
\end{tikzpicture}
\vspace*{-0.5cm}
\end{center}
\caption{Channel estimation MSE and resulting SNR degradation dependent on input SNR.}
\label{fig:CHMSE}
\end{figure}
\fi

In our channel model we assume that the signal arriving at each time instant consists of a single ray. We further assume that the direction of arrival is uniformly distributed and a \ac*{ULA} with element spacing of $\lambda/2$ is employed, so that the elements of the spatial correlation matrix can be calculated as:
\begin{equation}
	\begin{gathered}
	\left[\boldsymbol{R}_s\right]_{m1,m2} = E\left[e^{j\pi\sin(\theta)(m1-m2)}\right] \\= \frac{1}{2\pi}\int\limits^{\pi}_{-\pi} e^{j\pi\sin(\theta)(m1-m2)} d\theta.
	\end{gathered}
\end{equation}
This is the definition of the zeroth order Bessel function of the first kind
\begin{equation}
	\left[\boldsymbol{R}_s\right]_{m1,m2} = \text{J}_0(\pi(m1-m2)).
\end{equation}
It is important to mention that in the case of hybrid beamforming the spatial correlation after the analog combining is unknown. 
Since we select the beamforming vectors independently for each RF chain we assume that the resulting channels are spatially uncorrelated.  Thus, for this case the spatial correlation matrix $\boldsymbol{R}_s$ is an identity matrix. 
Based on this calculation we can also generate the spatial interpolation matrix $\boldsymbol{A}_s$ based on the Wiener filter equation as
\begin{equation}
	\boldsymbol{A}_s =\boldsymbol{R}_s (\boldsymbol{R}_s + \sigma_{\eta}^2\boldsymbol{I})^{-1}.
\end{equation}

Now we have assembled all the necessary mathematical tools to calculate the mean channel estimation error from the given reference signal pattern.
A maximum of four \glspl*{UE} are considered. For this system setup it is sufficient to generate reference sequence by cyclic shifting and multiplication with a orthogonal cover code of a Gold sequence
sequence as in the future 5G \ac*{NR} standard \cite{3gpp.38.211}. 
As shown in Fig. \ref{fig:RSPattern}, in contrast to \ac*{LTE}, the \ac*{DMRS} are located in separate \ac*{OFDM} symbols. 
As we can see from the figure the different \ac*{DMRS} groups are always allocated to adjacent \acp*{SC}. 
For the purpose of calculating the channel estimation mean square error we used the same channels statistics we use later for the rate calculation. Fig. \ref{fig:CHMSE} shows the calculated MSE 
and the corresponding \ac*{SNR} degradation. For the \ac*{SNR} degradation we assume a \ac*{SISO} system and that the channel estimation error is independent of the actual channel realizations. 

\subsection{Power Model}
For modeling the power of the different RF frontends we use the model described in \cite{KILIANJSACEE}.
This power model is based on components reported in the literature for the WiGig standard (802.11ad) operating in the 60 GHz \ac*{ISM} band. 
Since this standard was released in 2012 we can safely assume that the designs have reached sufficient maturity to represent low cost, low power 
power \ac*{CMOS} implementation. 
Table \ref{tab:powertab100} shows the power consumption of the different components.
\begin{table}
	\renewcommand{\arraystretch}{1.3}
	\caption{Components with power consumption.}
	\label{tab:powertab100}
	\centering
		\begin{tabularx}{0.95\columnwidth}{|X|X|X|}
			\hline
			label & component & power consumption \\ \hline \hline
			$P_{LO}$ & LO & 22.5 mW \\ \hline
			$P_{LNA}$ & LNA & 5.4 mW \\ \hline
			$P_{M}$ & mixer & 0.3 mW \\ \hline
			$P_{H}$ & 90$^{\circ}$ hybrid and LO buffer & 3 mW \\ \hline
			$P_{LA}$ & LA & 0.8 mW \\ \hline
			$P_{1}$ & 1-bit ADC & 0 mW \\ \hline
			$P_{PS}$ & phase shifter & 2 mW\\ \hline
			$P_{VGA}$ & VGA & 2 mW \\ \hline
			$P_{ADC}$ & ADC & $15~\mu \text{W/GHz}$ $\cdot f_s 2^{\text{ENOB}}$ \\ \hline
		\end{tabularx}
\end{table}

With the power consumption of the components, it is possible to compute the power consumption of the overall receiver front-end $P_R$ as:
\begin{equation}
	\begin{gathered}
		P_{R} =  P_{LO} + M_R\left(P_{LNA} + P_{H} + 2P_{M}\right) +\\
		\text{flag}_{C}\left(M_R P_{PS}\right)  +\\
		M_h\left(  \neg \text{flag}_{1\text{bit}}\left(2P_{VGA} + 2P_{ADC1}\right) + \text{flag}_{1\text{bit}}\left( 2P_{LA} \right)\right) +\\
		M_t\left(  \neg \text{flag}_{1\text{bit}}\left(2P_{VGA} + 2P_{ADC2}\right) + \text{flag}_{1\text{bit}}\left( 2P_{LA} \right)\right),
	\end{gathered}
\end{equation}
where $\text{flag}_{C}$ indicates if analog combining is used:
\begin{equation}
		\text{flag}_{C} = \left\{\begin{array}{ll} 0, & M_{RFE} =M_h + M_t =  M_R, M_C = 1 \\
         1, & \text{else}\end{array}\right. .
\end{equation}
The variable $\text{flag}_{1\text{bit}}$ indicates if 1 or higher resolution quantization is used. 
The operator $\neg$ represents a logic negation. 
In the case of 1-bit quantization, the power consumption of the \ac*{VGA} is replaced by that of the \ac*{LA} and the power consumption of the
1-bit quantizer is negligible compared to the rest of the front-end. This formula now contains all special cases of digital beamforming ($M_{RFE} = M_R$),
analog beamforming ($M_R > 0 ~\text{and}~ M_{RFE} = 1$) and hybrid beamforming.

\section{Rate Expression}
\subsection{Allocation of RF chains for hybrid beamforming with multiple users}
For the following calculations we assume that adjacent antennas are connected to one RF-chain.
Finding the optimal configuration of the phase shifters at each antenna to support $U$ users is a non-convex problem, which does not have a trivial solution. 
Thus, we introduce a number of simplifications that make the problem tractable. At the same time these simplifications are modeling the behavior of
practical beamforming systems like WiGig (802.11ad) \cite{WIGIGSTDORIGINAL, 80211ayBF}.

The overall procedure of selecting the beams is described in the following paragraph in an abstract way. Afterwards, the mathematical details are presented in the description of the algorithm. 
We limit the search for the optimal beamforming configuration in the following way: First, we search for the best beam for each user 
$i$ and RF-chain $j$ combination under the assumption that the other users are not present and record the corresponding receive power. Afterwards,
the RF-chains are allocated to the users in a resource-fair manner, starting from the RF chain and user with the highest receive power. 

As we showed in \cite{KILIANJSACEE}, if the receive antennas form a \ac*{ULA} at each subarray of $M_C$ elements and limiting the beams to receive the signal from only on spacial direction, we 
achieve 10\% error while having a codebook size of $4 M_C$. The first part of the algorithm is thus selecting the best beamforming vectors per UE. 
Since we assume that all subarrays have the same size $M_C$ we initialize the set of all possible directions $\mathbb{B}$ with $4 M_C$ values uniformly spaced from $-\pi$ to $\pi$:
\begin{equation}
	\mathbb{B} = \left\{ \phi_1, \phi_2, \cdots, \phi_{4M_{C}}\right\}, \phi_j = -\pi + \frac{j2\pi}{M_C}.
\end{equation}
Afterwards, for each user $u$ and each sub-array $i$, all direction are tested, and the one leading to the largest receive power and the corresponding index are stored
\begin{equation}
\begin{gathered}
				p(j) = \sum\limits_{l = 0}^{L-1}\left\vert\left\vert \boldsymbol{w}_{j}^H \boldsymbol{H}_u^i[l]\right\vert\right\vert^2_2 \\
				 [\boldsymbol{P}]_{u,i} = \max\limits_j p(j) \\
				 [\boldsymbol{J}]_{u,i} = \arg\max\limits_j p(j),
\end{gathered}
\end{equation}
with the vector ${w}_{j}$ defined as:
\begin{equation}
	\boldsymbol{w}_{j} = \left[1, e^{\phi_j}, \cdots, e^{(M_{C}-1)\phi_j} \right]^H.
\end{equation}
The matrices $\boldsymbol{P}$ and $\boldsymbol{J}$ contain the optimal power and the corresponding direction for all combinations of user $u$ and subarray $j$. 

The next step is to select which subarray should take which configuration. We at first fill the set $\mathbb{U}$ and $\mathbb{I}$ with all users and subarrays
\begin{equation}
	\mathbb{U} = \{1, \cdots, U\},~~\mathbb{I} = \{1, \cdots, M_{\text{RFE}}\}.
\end{equation}
Then we select the subarray-user combination leading to the largest receive power and allocate the array steering vector of the selected subarray to this configuration. 
Since this subarray and user are now allocated we remove them from the sets $\mathbb{U}$ and $\mathbb{I}$.
If the set of remaining users is empty we reset it to all possible users. 
This procedure is repeated until all subarrays are allocated. 
It ensures that
the subarrays are distributed among the users under a resource fair constraint. 
In addition the selection of the those with higher power also ensures that the rate
is optimized. It is important to mention that only selecting the RF-chains according to the ones providing the largest receive power, even if considered for all users
would lead to starvation of the users with the worst channels. Since this is not desirable we adopted the above procedure. The entire process is summarized in Algorithm \ref{alg:beamforming}.
\begin{figure}
\vspace*{-0.1cm}
\begin{algorithm}[H]
	\caption{Selection of the beamforming vectors.}
	\begin{algorithmic}[1]
		\Require{$\boldsymbol{H}[l]$, $U$, $M_{\text{RFE}}$ and $M_{C}$}
		\State $\mathbb{B} \gets \left\{\phi_1, \phi_2, \cdots, \phi_{4M_{C}}\right\}$
		\For{$u \gets 1 \textrm{ to } U$}
			\For{$i \gets 1 \textrm{ to } M_{\text{RFE}}$}
				\For{$j \gets 1 \textrm{ to } 4M_{C}$}
					\State $\boldsymbol{w}_{j} \gets \left[1, e^{\phi_j}, \cdots, e^{(M_{C}-1)\phi_j} \right]^H$
					\State $p(j) \gets \sum\limits_{l = 0}^{L-1}\left\vert\left\vert \boldsymbol{w}_{j}^H \boldsymbol{H}_u^i[l]\right\vert\right\vert^2_2 $
				\EndFor
				\State $ [\boldsymbol{P}]_{u,i} \gets \max\limits_j p(j)$
				\State $ [\boldsymbol{J}]_{u,i} \gets \arg\max\limits_j p(j)$
			\EndFor
		\EndFor
		\State $\mathbb{U} \gets \{1, \cdots, U\}$
		\State $\mathbb{I} \gets \{1, \cdots, M_{\text{RFE}}\}$
		\For{$i \gets 1 \textrm{ to } M_{\text{RFE}}$}
		\State $\hat{u}, \hat{i} \gets \arg\!\max\limits_{u \in \mathbb{U}, i \in \mathbb{I}} [\boldsymbol{P}]_{u,i}$
		\State $\hat{j} \gets [\boldsymbol{J}]_{\hat{u}, \hat{i}}$
		\State $ \boldsymbol{w}^{\hat{i}}_R \gets \left[1, e^{\phi_{\hat{j}}}, \cdots, e^{(M_{C}-1)\phi_{\hat{j}}} \right]^H$
		\State $\mathbb{I} \gets \mathbb{I}\setminus \hat{i}$
		\State $\mathbb{U} \gets \mathbb{U}\setminus \hat{u}$
		\If{$\mathbb{U} = \varnothing$} 
			\State $\mathbb{U} \gets \{1, \cdots, U\}$
		\EndIf
		\EndFor
		\State \Return $\boldsymbol{w}^i_R~\forall i = \{1, \hdots, M_{\text{RFE}}\}$
  \end{algorithmic}
  \label{alg:beamforming}
\end{algorithm}
\vspace*{-0.3cm}
\end{figure}

\subsection{Modeling the Quantization}
As in \cite{AQNMAMINE, HLOWADC}, we use the Bussgang theorem to decompose the signal after quantization in a signal component and an uncorrelated quantization error $\boldsymbol{e}$:
\begin{equation}
	\boldsymbol{r}[n] = Q(\boldsymbol{y}_{C}[n]) \approx \boldsymbol{F}\boldsymbol{y}_{C}[n] + \boldsymbol{e}[n],
\end{equation}
with $\boldsymbol{y}_{C}[n]$ being the signal after the analog combining at the receiver 
equal to $\boldsymbol{u}[n] + \boldsymbol{\eta}_r[n]$, where $\boldsymbol{u}[n]$ is the receive signal after the multipath channel. 
The operation $Q(\cdot)$ represents the quantization, which is performed separately for each element of the vector as well as their real and imaginary parts.
This includes the possibility of using \acp*{ADC} with different resolution at each element.

To include the quantization into the rate analysis we need to calculate $\boldsymbol{F}$ and the covariance matrix 
$\boldsymbol{R}_{\boldsymbol{e}\boldsymbol{e}}$of $\boldsymbol{e}[n]$. 
The description in Appendix \ref{appendixAQNM} shows how to calculate these matrices from the receive covariance matrix $\boldsymbol{R}_{\boldsymbol{y}_{C}\boldsymbol{y}_{C}}$ and the quantization functions. For the calculation of the receive covariance matrix we reuse the formulas we derived in \cite{KILIANJSACEE}. 
To simplify the notation we use the operands defined in Appendix \ref{appendixAQNM}
\begin{equation}
\begin{gathered}
	 \boldsymbol{F} = \text{TF}(Q^1(\cdot), \cdots, Q^{M_{\text{RFE}}}).\\
	 \boldsymbol{R}_{\boldsymbol{r}\boldsymbol{r}} = \text{T}\left(\boldsymbol{R}_{\boldsymbol{y}_C\boldsymbol{y}_C}, Q^1(\cdot), \cdots, Q^{M_{\text{RFE}}}\right).
\end{gathered}
\end{equation}
With these results we can calculate the quantization error covariance matrix as
\begin{equation}
	 \boldsymbol{R}_{\boldsymbol{e}\boldsymbol{e}} = \boldsymbol{R}_{\boldsymbol{r}\boldsymbol{r}} - \boldsymbol{F} \boldsymbol{R}_{\boldsymbol{y}_C\boldsymbol{y}_C} \boldsymbol{F}
\end{equation}

Now we can calculate the effective channel $\boldsymbol{H}^\prime[l]$ and noise covariance matrix  $\boldsymbol{R}_{\boldsymbol{\eta}^\prime\boldsymbol{\eta}^\prime}$of the overall system including the analog combing and the quantization:
\begin{equation}
	\boldsymbol{H}^\prime[l] = \boldsymbol{F}\boldsymbol{W}_R^H\boldsymbol{H}[l],
\end{equation}
and
\begin{equation}
	\boldsymbol{R}_{\boldsymbol{\eta}^\prime\boldsymbol{\eta}^\prime} =  \boldsymbol{F}\boldsymbol{W}_R^H \boldsymbol{R}_{\boldsymbol{\eta}^\prime_R\boldsymbol{\eta}^\prime_R} \boldsymbol{W}_R \boldsymbol{F}^H +
	\boldsymbol{R}_{\boldsymbol{e}\boldsymbol{e}}.
	\label{eq:Rnndash}
\end{equation}

It is also important to mention that many previous evaluations (\cite{AQNMAMINE, QINGOPAQNM, HLOWADC, ACHCDCCOMP, ONERLOWRES}) only use a diagonal approximation of the quantization error covariance matrix. As we show in \cite{KILIANJSACEE}, including the off-diagonal elements in the evaluation can have a dramatic impact on the overall performance. Therefore, we generalized our previously derived formulas for the case with different quantization functions to also include the off-diagonal elements in this evaluation. 

\subsection{Modeling the Channel Estimation Error}
After the model for the transmit impairments, the analog combining and the quantization error we have a set of equations that looks fairly similar to a standard MIMO system.
We chose to model the channel estimation error as additional noise independent of receive channel.
This is different from the work in \cite{YOO2006}. 
In this work the channel estimation error is also modeled as additional noise. But in addition the useful signal power is divided between the estimated channel and the channel estimation noise. 
This has the effect that for cases leading to a large estimation error, the resulting signal receive power goes and thus the rate go to zero.
If we look at our simulation of the channel estimation error in Fig. \ref{fig:CHMSE} this would be the case for the very low \ac*{SNR} range from -30 to -10 dB. 
This contradicts the practical observation, that communication at a \ac*{SNR} as low as -10 dB for a \ac*{SISO} system is possible \cite{Ratasuk2016}.
For a practical massive \ac*{MIMO} system this would mean that regardless of the number of antennas it is not possible to be used at low \ac*{SNR}.
We therefore think that modeling the channel estimation error as noise is more suitable to reflect the behavior of a practical system.

The overall covariance matrix of the channel estimation error $\boldsymbol{R}_{\boldsymbol{w}\boldsymbol{w}}$ is defined as a sum of the per user $\boldsymbol{R}_{\boldsymbol{w}_u\boldsymbol{w}_u}$
\begin{equation}
	\boldsymbol{R}_{\boldsymbol{w}[f]\boldsymbol{w}[f]} = \sum\limits^{U}_{u=1}\boldsymbol{R}_{\boldsymbol{w}_u[f]\boldsymbol{w}_u[f]},
\end{equation}
where the variance of each element of $\boldsymbol{R}_{\boldsymbol{w}_u[f]\boldsymbol{w}_u[f]}$ depends on the channel estimation error $\sigma^2_u$ and the
actual power of the channel at the corresponding frequency bin $f$ on antenna $m$:
\begin{equation}
	\left[\boldsymbol{R}_{\boldsymbol{w}_u[f]\boldsymbol{w}_u[f]}\right]_{m,m} = \vert [\boldsymbol{h}_u[f]]_m \vert^2 \sigma^2_u.
	\end{equation}
We model each matrix $\boldsymbol{R}_{\boldsymbol{w}_u[f]\boldsymbol{w}_u[f]}$ to be spatially white and thus a diagonal matrix. 
The values $\sigma^2_u$ are determined by calculating the average \ac*{SNR} per antenna per user and then obtaining the corresponding MSE $\sigma^2_u$
from the simulation shown in Fig. \ref{fig:CHMSE}. 

We combine the this calculation into the operator $\text{TE}(\cdot)$
\begin{equation}
	\boldsymbol{R}_{\boldsymbol{w}[f]\boldsymbol{w}[f]} = \text{TE}(\boldsymbol{H}[f], \boldsymbol{R}_ {\boldsymbol{y}\boldsymbol{y}}, \boldsymbol{R}_ {\boldsymbol{\eta}\boldsymbol{\eta}}).
\end{equation}

\subsection{Combined Rate Expression}
At this point we have all the necessary information to calculate the sum rate for the given scenario.
We make a number of approximations that make the expression tractable:
\begin{itemize}
	\item Assume $\boldsymbol{x}(f)$ is Gaussian
	\item $\boldsymbol{w}_R^i$ are selected from the derived finite set separately for each antenna group based on an \ac*{SNR} criteria
	\item Quantization is modeled as additive Gaussian noise with the \ac*{AQNM} model including the off-diagonal elements
	\item No collaboration among the users
\end{itemize}
With these simplifications the $\boldsymbol{w}_R^i$ are already defined and we can transform the problem into a frequency domain equation.

\begin{figure}
\vspace*{-0.1cm}
\begin{algorithm}[H]
	\caption{Combined multipath channel from each user $\boldsymbol{H}[l]$, combined transmit impairments \ac*{EVM} co-variance matrix 
	$\boldsymbol{R}_{\boldsymbol{\eta_T}\boldsymbol{\eta_T}}$, combined maximum transmit power constraint $P_{Tx}\boldsymbol{I}$,
	receiver noise covariance matrix $\boldsymbol{R}_{\boldsymbol{\eta}\boldsymbol{\eta}}$, frequency band from $f_1$ to $f_2$,
	quantization function $Q^m(\cdot)$ separate for each receiver chain $m$ and channel statistics and number of frequency bins $N_f$.}
	\begin{algorithmic}
    		\Require{$\boldsymbol{R}_{\boldsymbol{\eta_T}\boldsymbol{\eta_T}}$, 
    		$\boldsymbol{R}_{\boldsymbol{\eta}\boldsymbol{\eta}}$, $\boldsymbol{H}[l]$,  $P_{Tx}$, $f_1$, $f_2$ and $Q_b(\cdot)$}
		\State $\boldsymbol{H}[f] \gets \mathcal{F}(\boldsymbol{H}[l])$
		\State $\boldsymbol{R}_{\boldsymbol{x}[f]\boldsymbol{x}[f]} \gets P_{Tx}\boldsymbol{I}$
		\State $\boldsymbol{R}_{\boldsymbol{y}\boldsymbol{y}} \gets \sum\limits_{f_1}^{f_2}\boldsymbol{H}[f] \left( \boldsymbol{R}_{\boldsymbol{x}(f)\boldsymbol{x}[f]} + \boldsymbol{R}_{\boldsymbol{\eta_T}\boldsymbol{\eta_T}} \right)\boldsymbol{H}^H[f] + \boldsymbol{R}_{\boldsymbol{\eta}\boldsymbol{\eta}}$
		\State $\boldsymbol{R}_{\boldsymbol{r}\boldsymbol{r}} \gets \text{T}(\boldsymbol{R}_{\boldsymbol{y}\boldsymbol{y}}, Q^m(\cdot), \cdots, Q^{M_{RFE}}(\cdot))$
		\State $\boldsymbol{F} \gets \text{TF}(\boldsymbol{R}_{\boldsymbol{y}\boldsymbol{y}}, Q^m(\cdot), \cdots, Q^{M_{RFE}}(\cdot))$
		\State $\boldsymbol{R}_{\boldsymbol{\eta}^\prime\boldsymbol{\eta}^\prime} \gets  \boldsymbol{F}\boldsymbol{W}_R^H \boldsymbol{R}_{\boldsymbol{\eta}^\prime_R\boldsymbol{\eta}^\prime_R} \boldsymbol{W}_R \boldsymbol{F}^H + \boldsymbol{R}_{\boldsymbol{r}\boldsymbol{r}} - \boldsymbol{F} \boldsymbol{R}_{\boldsymbol{y}\boldsymbol{y}} \boldsymbol{F}$
		\State{$\boldsymbol{H}^\prime[l] \gets \boldsymbol{F}\boldsymbol{H}[l]~\forall l \in \{0, \hdots, L-1\}$}
		\State $\boldsymbol{H}^\prime[f] \gets \mathcal{F}(\boldsymbol{H}^\prime[l])$
		\State $\boldsymbol{R}_{\boldsymbol{w}[f]\boldsymbol{w}[f]} \gets \text{TE}(\boldsymbol{H}[f], \boldsymbol{R}_ {\boldsymbol{y}\boldsymbol{y}}, \boldsymbol{R}_ {\boldsymbol{\eta}\boldsymbol{\eta}})~\forall f \in \left[f_1, f_2\right]$
		\State $\boldsymbol{R}_{\boldsymbol{\eta}^\prime[f]\boldsymbol{\eta}^\prime[f]} \gets \boldsymbol{R}_{\boldsymbol{\eta}^\prime\boldsymbol{\eta}^\prime} + \boldsymbol{R}_{\boldsymbol{w}[f]\boldsymbol{w}[f]}~\forall f \in \left[f_1, f_2\right]$
		\State{$\boldsymbol{A}[f] \gets  \boldsymbol{I} + \boldsymbol{R}^{-1}_{\boldsymbol{\eta}^\prime[f]\boldsymbol{\eta}^\prime[f]} (\boldsymbol{H}^\prime[f]) \boldsymbol{R}_{\boldsymbol{x}[f]\boldsymbol{x}[f]} (\boldsymbol{H}^{\prime}[f])^H ~ \forall f \in \left[f_1, f_2\right]$}
		\State $R = \frac{1}{N_f}\sum\limits^{f_2}_{f_1} \log_2\left(\text{det}\left(\boldsymbol{A}[f]\right)\right)$
		\State \Return $R$
  \end{algorithmic}
  \label{alg:RateQuantEstErr}
\end{algorithm}
\vspace*{-0.3cm}
\end{figure}
\ifCLASSOPTIONdraftcls
\else
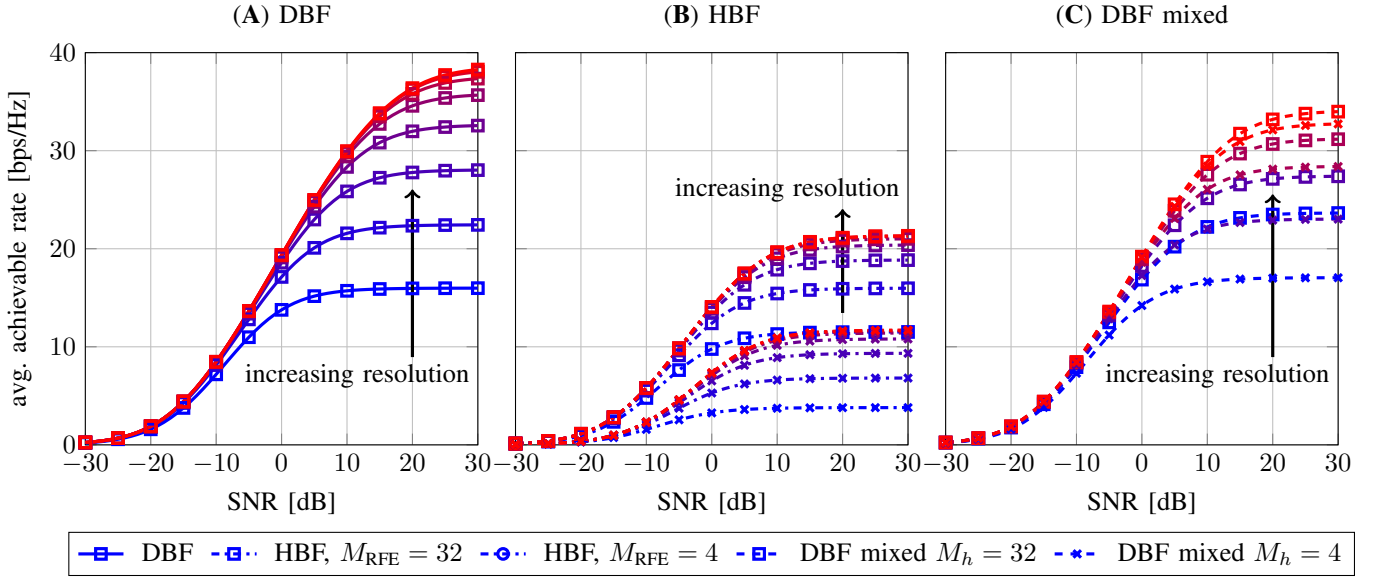
\begin{figure*}[!t]
\begin{center}
\normalsize
\begin{tikzpicture}
\begin{groupplot}[group style={group size= 3 by 1, horizontal sep=0.5cm}, width=18.1cm,
]
    \nextgroupplot[title={(\textbf{A}) DBF}, width=6.8cm, height=6.8cm, ylabel={avg. achievable rate [bps/Hz]}, grid,  legend cell align=left, ymin=0, ymax=40, xtick distance=10,
                  		xlabel={SNR [dB]}, xmin =-30, xmax=30, legend pos=north west, legend cell align=left, legend columns=1]
                  		
          	\addplot[very thick, blue, mark=square, mark repeat*=5] table[x = x, y=y]{./SimulationResults/PlotData/results_MAC4_Multipath_1_64_CEErr__DBF_bits_1_DBF_SNR_SE.txt};
		\addplot[very thick, BTORMIX16, mark=square, mark repeat*=5] table[x = x, y=y]{./SimulationResults/PlotData/results_MAC4_Multipath_1_64_CEErr__DBF_bits_2_DBF_SNR_SE.txt};
		\addplot[very thick, BTORMIX26, mark=square, mark repeat*=5] table[x = x, y=y]{./SimulationResults/PlotData/results_MAC4_Multipath_1_64_CEErr__DBF_bits_3_DBF_SNR_SE.txt};
		\addplot[very thick, BTORMIX36, mark=square, mark repeat*=5] table[x = x, y=y]{./SimulationResults/PlotData/results_MAC4_Multipath_1_64_CEErr__DBF_bits_4_DBF_SNR_SE.txt};
		\addplot[very thick, BTORMIX46, mark=square, mark repeat*=5] table[x = x, y=y]{./SimulationResults/PlotData/results_MAC4_Multipath_1_64_CEErr__DBF_bits_5_DBF_SNR_SE.txt};
		\addplot[very thick, BTORMIX56, mark=square, mark repeat*=5] table[x = x, y=y]{./SimulationResults/PlotData/results_MAC4_Multipath_1_64_CEErr__DBF_bits_6_DBF_SNR_SE.txt};
		\addplot[very thick, BTORMIX66, mark=square, mark repeat*=5] table[x = x, y=y]{./SimulationResults/PlotData/results_MAC4_Multipath_1_64_CEErr__DBF_bits_7_DBF_SNR_SE.txt};
		\addplot[very thick, red, mark=square, mark repeat*=5] table[x = x, y=y]{./SimulationResults/PlotData/results_MAC4_Multipath_1_64_CEErr__DBF_bits_8_DBF_SNR_SE.txt};
		
		\coordinate (c1) at (rel axis cs:0,1);
		\node[anchor=west]  at (axis cs:-7,7){increasing resolution};
   		\node (source) at (axis cs:20,8){};  		
   		\node (destination) at (axis cs:20,27){};
   		\draw[->, very thick](source)--(destination);
    \nextgroupplot[title={(\textbf{B}) HBF},width=6.8cm, height=6.8cm, grid,  legend cell align=left, ymin=0, ymax=40, yticklabels={,,}, xtick distance=10,
                  		xlabel={SNR [dB]}, xmin =-30, xmax=30, legend pos=north west, legend cell align=left, legend columns=1]
 			
          	\addplot[very thick, dash dot, blue, mark=square, mark repeat*=5, mark options={solid}] table[x = x, y=y]{./SimulationResults/PlotData/results_MAC4_Multipath_1_64_CEErr_HBF_MC_2_bits_1_DBF_SNR_SE.txt};
          	\addplot[very thick, dash dot, BTORMIX16, mark=square, mark repeat*=5, mark options={solid}] table[x = x, y=y]{./SimulationResults/PlotData/results_MAC4_Multipath_1_64_CEErr_HBF_MC_2_bits_2_DBF_SNR_SE.txt};
          	\addplot[very thick, dash dot, BTORMIX26, mark=square, mark repeat*=5, mark options={solid}] table[x = x, y=y]{./SimulationResults/PlotData/results_MAC4_Multipath_1_64_CEErr_HBF_MC_2_bits_3_DBF_SNR_SE.txt};
          	\addplot[very thick, dash dot, BTORMIX36, mark=square, mark repeat*=5, mark options={solid}] table[x = x, y=y]{./SimulationResults/PlotData/results_MAC4_Multipath_1_64_CEErr_HBF_MC_2_bits_4_DBF_SNR_SE.txt};
          	\addplot[very thick, dash dot, BTORMIX46, mark=square, mark repeat*=5, mark options={solid}] table[x = x, y=y]{./SimulationResults/PlotData/results_MAC4_Multipath_1_64_CEErr_HBF_MC_2_bits_5_DBF_SNR_SE.txt};
          	\addplot[very thick, dash dot, BTORMIX56, mark=square, mark repeat*=5, mark options={solid}] table[x = x, y=y]{./SimulationResults/PlotData/results_MAC4_Multipath_1_64_CEErr_HBF_MC_2_bits_6_DBF_SNR_SE.txt};
          	\addplot[very thick, dash dot, BTORMIX66, mark=square, mark repeat*=5, mark options={solid}] table[x = x, y=y]{./SimulationResults/PlotData/results_MAC4_Multipath_1_64_CEErr_HBF_MC_2_bits_7_DBF_SNR_SE.txt};
          	\addplot[very thick, dash dot, red, mark=square, mark repeat*=5, mark options={solid}] table[x = x, y=y]{./SimulationResults/PlotData/results_MAC4_Multipath_1_64_CEErr_HBF_MC_2_bits_8_DBF_SNR_SE.txt};

          	\addplot[very thick, blue, dash dot, mark=x, mark repeat*=5, mark options={solid}] table[x = x, y=y]{./SimulationResults/PlotData/results_MAC4_Multipath_1_64_CEErr_HBF_MC_16_bits_1_DBF_SNR_SE.txt};
          	\addplot[very thick, BTORMIX16, dash dot, mark=x, mark repeat*=5, mark options={solid}] table[x = x, y=y]{./SimulationResults/PlotData/results_MAC4_Multipath_1_64_CEErr_HBF_MC_16_bits_2_DBF_SNR_SE.txt};
          	\addplot[very thick, BTORMIX26, dash dot, mark=x, mark repeat*=5, mark options={solid}] table[x = x, y=y]{./SimulationResults/PlotData/results_MAC4_Multipath_1_64_CEErr_HBF_MC_16_bits_3_DBF_SNR_SE.txt};
          	\addplot[very thick, BTORMIX36, dash dot, mark=x, mark repeat*=5, mark options={solid}] table[x = x, y=y]{./SimulationResults/PlotData/results_MAC4_Multipath_1_64_CEErr_HBF_MC_16_bits_4_DBF_SNR_SE.txt};
          	\addplot[very thick, BTORMIX46, dash dot, mark=x, mark repeat*=5, mark options={solid}] table[x = x, y=y]{./SimulationResults/PlotData/results_MAC4_Multipath_1_64_CEErr_HBF_MC_16_bits_5_DBF_SNR_SE.txt};
          	\addplot[very thick, BTORMIX56, dash dot, mark=x, mark repeat*=5, mark options={solid}] table[x = x, y=y]{./SimulationResults/PlotData/results_MAC4_Multipath_1_64_CEErr_HBF_MC_16_bits_6_DBF_SNR_SE.txt};
          	\addplot[very thick, BTORMIX66, dash dot, mark=x, mark repeat*=5, mark options={solid}] table[x = x, y=y]{./SimulationResults/PlotData/results_MAC4_Multipath_1_64_CEErr_HBF_MC_16_bits_7_DBF_SNR_SE.txt};
          	\addplot[very thick, red, dash dot, mark=x, mark repeat*=5, mark options={solid}] table[x = x, y=y]{./SimulationResults/PlotData/results_MAC4_Multipath_1_64_CEErr_HBF_MC_16_bits_8_DBF_SNR_SE.txt};
          	
          	\node[anchor=west]  at (axis cs:-7,26){increasing resolution};
   		\node (source) at (axis cs:20,12.5){};  		
   		\node (destination) at (axis cs:20,25){};
   		\draw[->, very thick](source)--(destination);
   		
    \nextgroupplot[title={(\textbf{C}) DBF mixed}, width=6.8cm, height=6.8cm, grid,  legend cell align=left, ymin=0, ymax=40, yticklabels={,,}, xtick distance=10,
                  		xlabel={SNR [dB]}, xmin =-30, xmax=30, legend to name=grouplegend1, legend style={legend columns=5,fill=none,draw=black,anchor=center,align=left, column sep=0.15cm},]
                  		
             	\addlegendimage{very thick, blue, mark=square}
 			\addlegendentry{DBF}   	
             	\addlegendimage{very thick, dash dot, blue, mark=square, mark options={solid}}
 			\addlegendentry{HBF, $M_{\text{RFE}} = 32$}   
 			\addlegendimage{very thick, blue, dash dot, mark=o, mark options={solid}}
 			\addlegendentry{HBF, $M_{\text{RFE}} = 4$}   	
             	\addlegendimage{very thick, blue, mark=square, mark options={solid}, dashed}
 			\addlegendentry{DBF mixed $M_h = 32$}   
 			\addlegendimage{very thick, blue, dashed, mark=x, mark options={solid}}
			\addlegendentry{DBF mixed $M_h = 4$}   	
 			
          	\addplot[very thick, blue, mark=square, mark repeat*=5, mark options={solid}, dashed] table[x = x, y=y]{./SimulationResults/PlotData/results_MAC4_Multipath_1_64_CEErr_32_32_5bit__DBF_bits_1_DBF_SNR_SE.txt};
          	\addplot[very thick, BTORMIX12, mark=square, mark repeat*=5, mark options={solid}, dashed] table[x = x, y=y]{./SimulationResults/PlotData/results_MAC4_Multipath_1_64_CEErr_32_32_5bit__DBF_bits_2_DBF_SNR_SE.txt};
          	\addplot[very thick, BTORMIX22, mark=square, mark repeat*=5, mark options={solid}, dashed] table[x = x, y=y]{./SimulationResults/PlotData/results_MAC4_Multipath_1_64_CEErr_32_32_5bit__DBF_bits_3_DBF_SNR_SE.txt};
          	\addplot[very thick, red, mark=square, mark repeat*=5, mark options={solid}, dashed] table[x = x, y=y]{./SimulationResults/PlotData/results_MAC4_Multipath_1_64_CEErr_32_32_5bit__DBF_bits_4_DBF_SNR_SE.txt};
          	
          	\addplot[very thick, blue, dashed, mark=x, mark repeat*=5, mark options={solid}] table[x = x, y=y]{./SimulationResults/PlotData/results_MAC4_Multipath_1_64_CEErr_60_4_5bit__DBF_bits_1_DBF_SNR_SE.txt};
          	\addplot[very thick, BTORMIX12, dashed, mark=x, mark repeat*=5, mark options={solid}] table[x = x, y=y]{./SimulationResults/PlotData/results_MAC4_Multipath_1_64_CEErr_60_4_5bit__DBF_bits_2_DBF_SNR_SE.txt};
          	\addplot[very thick, BTORMIX22, dashed, mark=x, mark repeat*=5, mark options={solid}] table[x = x, y=y]{./SimulationResults/PlotData/results_MAC4_Multipath_1_64_CEErr_60_4_5bit__DBF_bits_3_DBF_SNR_SE.txt};
          	\addplot[very thick, red, dashed, mark=x, mark repeat*=5, mark options={solid}] table[x = x, y=y]{./SimulationResults/PlotData/results_MAC4_Multipath_1_64_CEErr_60_4_5bit__DBF_bits_4_DBF_SNR_SE.txt};
          	
          	\coordinate (c2) at (rel axis cs:1,1);
          	\node[anchor=west]  at (axis cs:-7,7){increasing resolution};
   		\node (source) at (axis cs:20,8){};  		
   		\node (destination) at (axis cs:20,26.5){};
   		\draw[->, very thick](source)--(destination);
   		
\end{groupplot}
    	\coordinate (c3) at ($(c1)!.5!(c2)$);
    	\node[below] at (c3 |- current bounding box.south)
	{\ref{grouplegend1}};
\end{tikzpicture}
\vspace*{-0.8cm}
\caption{DBF, HBF and DBF mixed average achievable rate for $M_R = 64$, $U = 4$, $M_{\text{RFE}} \in \{ 4, 32\}$, $M_{h} \in \{4, 32\}$ and ADC resolution $b \in \{1, \cdots, 8\}$.}
\label{fig:SNRSECombined}
\vspace*{-0.3cm}
\hrulefill
\end{center}
\vspace*{-0.8cm}
\end{figure*}
\fi

The rate analysis is carried out for each frequency bin $f$ separately:
\begin{equation}
	\begin{gathered}
		R \le \int\displaylimits_{f1}^{f2}\max_{\boldsymbol{R}_{\boldsymbol{x}(f)\boldsymbol{x}(f)}}I(\boldsymbol{x}(f), \boldsymbol{r}(f)\vert\boldsymbol{H}^\prime(f))df \\
		\text{s.t.}~~\mathbb{E}[\vert\vert \boldsymbol{x}(f) \vert\vert^2_2] \le P_{Tx} ~\forall f \in [f_1, f_2],
	\end{gathered}
	\label{eq:achievrate}
\end{equation}
where $\boldsymbol{x}(f)$, $\boldsymbol{r}(f)$ and $\boldsymbol{H}^\prime(f)$ represent the input/output signal and equivalent channel of frequency bin $f$, and $I(\cdot)$ is the mutual information.
The frequencies $f_1$ and $f_2$ mark the borders of the band of interest in the equivalent baseband channel.
If the entire band covered by the sampling rate is not available to the system, the
parameters $f_1$ and $f_2$ have to account for the oversampling. 

Since all signals are represented by Gaussian random variables, we get the following expression for the mutual information:
\begin{equation}
	\begin{gathered}
		 I(\boldsymbol{x}(f), \boldsymbol{r}(f)\vert\boldsymbol{H}^\prime(f)) = \\ 
		\log_2\left(\text{det}\left( \boldsymbol{I} + \boldsymbol{R}^{-1}_{\boldsymbol{\eta}^\prime\boldsymbol{\eta}^\prime} \boldsymbol{H}^\prime(f) \boldsymbol{R}_{\boldsymbol{x}(f)\boldsymbol{x}(f)} \boldsymbol{H}^{\prime H}(f)\right)\right). \\
	\end{gathered}
\end{equation}
Due to the transmit noise, the modeling of the quantization and the channel estimation the effective noise 
covariance matrix $\boldsymbol{R}_{\boldsymbol{\eta}^\prime\boldsymbol{\eta}^\prime}$ and the effective channel 
$ \boldsymbol{H}^\prime(f)$ are dependent on the input covariance matrix $\boldsymbol{R}_{\boldsymbol{x}(f)\boldsymbol{x}(f)}$

The procedure of calculating the sum rate is summarized in algorithm \ref{alg:RateQuantEstErr}.

\section{Simulation Results}
Here we describe the chosen evaluation setup and the corresponding results. A basestation with 64 antennas ($M_R = 64$) receives the signal from 4 users ($U = 4$).
For the channel model of each user, identical modeling parameters but different realizations are chosen. 
We used the following parameters: $L = 128$, $P = 32$, $\beta = 0.5$. For the \ac*{HBF} system, $M_{\text{RFE}} \in \{4, 8, 16, 32\}$ \ac*{RF} chains are used. For \ac*{DBF} and \ac*{HBF} with uniform quantization we use a resolution of $b \in \{1, 2, 3, 4, 5, 6, 7, 8\}$ bits. 
For the case of \ac*{DBF} with mixed resolution \acp*{ADC} we used $M_h \in \{4, 8, 16, 32\}$ for the number of \acp*{ADC} with high resolution. The transmit power for all users is the same. Since on average the channel gain is the same the powers received from different users is similar. 
Since for the results with uniform quantization we found that the spectral efficiency at high \ac*{SNR} is maximized by an \ac*{ADC} resolution of 5 bits we chose $b_h = 5$. 
The resolution of the lower resolution \ac*{ADC} is chosen to be $b_l \in \{1, 2, 3, 4\}$ bits. 

\subsection{Average Achievable Rate Results}
Fig.
\ifCLASSOPTIONdraftcls
\ref{fig:SNRSEDBF} to \ref{fig:SNRSEDBFMixed}
\else
\ref{fig:SNRSECombined} (\textbf{A}) to (\textbf{C}) 
\fi
show the average achievable rate over 30 channel realizations. The resolution in bits increases from the top to bottom for each group of curves. 
From the \ac*{DBF} results in Fig.
\ifCLASSOPTIONdraftcls
\ref{fig:SNRSEDBF}
\else
\ref{fig:SNRSECombined} (\textbf{A})
\fi
we see that at high \ac*{SNR} the rate saturates and there is only minor improvement above a resolution of 5 bits. The reason for the saturation at high \ac*{SNR} are the transmitter impairments and channel estimation error. The same holds true for the \ac*{HBF} case in Fig.
\ifCLASSOPTIONdraftcls
\ref{fig:SNRSEHBF}
\else
\ref{fig:SNRSECombined} (\textbf{B})
\fi
. But due to the limited degrees of freedom the average achievable rate saturates at a lower value than the \ac*{DBF} case.
The results of the \ac*{DBF} mixed case in Fig.
\ifCLASSOPTIONdraftcls
\ref{fig:SNRSEDBFMixed}
\else
\ref{fig:SNRSECombined} (\textbf{C})
\fi
show that this approach can offer all possible rates in between the results of having only one \ac*{ADC} resolution, offering all possible values of energy and spectral efficiency around the values for \ac*{DBF} with only one \ac*{ADC} resolution.
Combining the observations of the achievable rate we can predict that the energy efficiency for an \ac*{ADC} resolution above 5 bits will not improve, since the achievable rate only shows limited improvement, while the power consumption of the 
front-end will dramatically increase. 

\subsection{Energy Efficiency Results}
We define the energy efficiency as the average achievable sum rate $R$ divided by the power consumption of the RF front-end $P_R$
\begin{equation}
 	\text{energy efficiency} = \frac{R}{P_R}.
\end{equation}
The scenarios in Fig.
\ifCLASSOPTIONdraftcls
\ref{fig:SEEE-15} to \ref{fig:SEEEmixed15}
\else
\ref{fig:SEEECombined} (\textbf{A}) to (\textbf{D})
\fi
show the achievable rate and energy efficiency for different \acp*{SNR}. For each curve the \ac*{ADC} resolution increases from the leftmost point of the curve. This point represents 1 bit resolution for all \acp*{ADC} or 1 bit resolution for the ones with 
lower resolution in the case of mixed-\ac*{ADC} \ac*{DBF}.
For all cases we see that the \ac*{DBF} system is more energy efficient compared to \ac*{HBF}.
The major reason for this is that the digital system retains all available degrees of freedom.
We can see that as the \ac*{SNR} increases (Fig.
\ifCLASSOPTIONdraftcls
\ref{fig:SEEE-15} to \ref{fig:SEEE15}
\else
\ref{fig:SEEECombined} (\textbf{A}) to (\textbf{C})
\fi
) the smaller the improvement of additional \ac*{RF} chains. The explanation for this is that even though we gain more degrees of freedom we still need to divide them among the users.
In Fig.
\ifCLASSOPTIONdraftcls
\ref{fig:SEEE15}
\else
\ref{fig:SEEECombined} (\textbf{C})
\fi
we see that there is little difference between having 8 or 16 \ac*{RF} chains. 

As the \ac*{SNR} increases from 
\ifCLASSOPTIONdraftcls
Fig. \ref{fig:SEEE-15} to \ref{fig:SEEE15}
\else
Fig. \ref{fig:SEEECombined} (\textbf{A}) to (\textbf{C})
\fi
the optimal resolution in terms of energy efficiency improves. 
As predicted from the achievable rate curves, above a resolution of 5 bits the energy efficiency decreases for all cases.
The results for \ac*{DBF} with mixed configurations in Fig. 
\ifCLASSOPTIONdraftcls
\ref{fig:SEEEmixed15}
\else
\ref{fig:SEEECombined} (\textbf{A}) to (\textbf{D})
\fi
show that these curves are tightly clustered around the curves for the case with only one resolution. This shows that this approach can achieve all possible different values in the rate - energy efficiency trade-off. 

\ifCLASSOPTIONdraftcls
\begin{figure}
\centering
\begin{tikzpicture}
    \begin{axis}[width=0.95*8.8cm, height=0.95*8.8cm, ylabel={avg. achievable rate [bps/Hz]}, grid,  legend cell align=left, ymin=0, ymax=40,
                  		xlabel={SNR [dB]}, xmin =-30, xmax=30, legend pos=north west, legend cell align=left, legend columns=1]
                  	
             \addlegendimage{very thick, blue, mark=square}
 		\addlegendentry{DBF}   		
                  		
          	\addplot[very thick, blue, mark=square, mark repeat*=5] table[x = x, y=y]{./SimulationResults/PlotData/results_MAC4_Multipath_1_64_CEErr__DBF_bits_1_DBF_SNR_SE.txt};
		\addplot[very thick, BTORMIX16, mark=square, mark repeat*=5] table[x = x, y=y]{./SimulationResults/PlotData/results_MAC4_Multipath_1_64_CEErr__DBF_bits_2_DBF_SNR_SE.txt};
		\addplot[very thick, BTORMIX26, mark=square, mark repeat*=5] table[x = x, y=y]{./SimulationResults/PlotData/results_MAC4_Multipath_1_64_CEErr__DBF_bits_3_DBF_SNR_SE.txt};
		\addplot[very thick, BTORMIX36, mark=square, mark repeat*=5] table[x = x, y=y]{./SimulationResults/PlotData/results_MAC4_Multipath_1_64_CEErr__DBF_bits_4_DBF_SNR_SE.txt};
		\addplot[very thick, BTORMIX46, mark=square, mark repeat*=5] table[x = x, y=y]{./SimulationResults/PlotData/results_MAC4_Multipath_1_64_CEErr__DBF_bits_5_DBF_SNR_SE.txt};
		\addplot[very thick, BTORMIX56, mark=square, mark repeat*=5] table[x = x, y=y]{./SimulationResults/PlotData/results_MAC4_Multipath_1_64_CEErr__DBF_bits_6_DBF_SNR_SE.txt};
		\addplot[very thick, BTORMIX66, mark=square, mark repeat*=5] table[x = x, y=y]{./SimulationResults/PlotData/results_MAC4_Multipath_1_64_CEErr__DBF_bits_7_DBF_SNR_SE.txt};
		\addplot[very thick, red, mark=square, mark repeat*=5] table[x = x, y=y]{./SimulationResults/PlotData/results_MAC4_Multipath_1_64_CEErr__DBF_bits_8_DBF_SNR_SE.txt};
		
		\node[anchor=west]  at (axis cs:-5,9){increasing resolution};
   		\node (source) at (axis cs:20,10){};  		
   		\node (destination) at (axis cs:20,25){};
   		\draw[->, very thick](source)--(destination);
     \end{axis}
\end{tikzpicture}
\vspace*{-0.5cm}
\caption{DBF average achievable rate for $M_R = 64$, $U = 4$ and ADC resolution $b \in \{1, \cdots, 8\}$.}
\label{fig:SNRSEDBF}
\end{figure}
\begin{figure}
\centering
\begin{tikzpicture}
    \begin{axis}[width=0.95*8.8cm, height=0.95*8.8cm, ylabel={avg. achievable rate [bps/Hz]}, grid,  legend cell align=left, ymin=0, ymax=40,
                  		xlabel={SNR [dB]}, xmin =-30, xmax=30, legend pos=north west, legend cell align=left, legend columns=1]
                  		
             	\addlegendimage{very thick, dash dot, blue, mark=square, mark options={solid}}
 			\addlegendentry{HBF, $M_{\text{RFE}} = 32$}   
 			\addlegendimage{very thick, blue, dash dot, mark=o, mark options={solid}}
 			\addlegendentry{HBF, $M_{\text{RFE}} = 4$}   	
 			
          	\addplot[very thick, dash dot, blue, mark=square, mark repeat*=5, mark options={solid}] table[x = x, y=y]{./SimulationResults/PlotData/results_MAC4_Multipath_1_64_CEErr_HBF_MC_2_bits_1_DBF_SNR_SE.txt};
          	\addplot[very thick, dash dot, BTORMIX16, mark=square, mark repeat*=5, mark options={solid}] table[x = x, y=y]{./SimulationResults/PlotData/results_MAC4_Multipath_1_64_CEErr_HBF_MC_2_bits_2_DBF_SNR_SE.txt};
          	\addplot[very thick, dash dot, BTORMIX26, mark=square, mark repeat*=5, mark options={solid}] table[x = x, y=y]{./SimulationResults/PlotData/results_MAC4_Multipath_1_64_CEErr_HBF_MC_2_bits_3_DBF_SNR_SE.txt};
          	\addplot[very thick, dash dot, BTORMIX36, mark=square, mark repeat*=5, mark options={solid}] table[x = x, y=y]{./SimulationResults/PlotData/results_MAC4_Multipath_1_64_CEErr_HBF_MC_2_bits_4_DBF_SNR_SE.txt};
          	\addplot[very thick, dash dot, BTORMIX46, mark=square, mark repeat*=5, mark options={solid}] table[x = x, y=y]{./SimulationResults/PlotData/results_MAC4_Multipath_1_64_CEErr_HBF_MC_2_bits_5_DBF_SNR_SE.txt};
          	\addplot[very thick, dash dot, BTORMIX56, mark=square, mark repeat*=5, mark options={solid}] table[x = x, y=y]{./SimulationResults/PlotData/results_MAC4_Multipath_1_64_CEErr_HBF_MC_2_bits_6_DBF_SNR_SE.txt};
          	\addplot[very thick, dash dot, BTORMIX66, mark=square, mark repeat*=5, mark options={solid}] table[x = x, y=y]{./SimulationResults/PlotData/results_MAC4_Multipath_1_64_CEErr_HBF_MC_2_bits_7_DBF_SNR_SE.txt};
          	\addplot[very thick, dash dot, red, mark=square, mark repeat*=5, mark options={solid}] table[x = x, y=y]{./SimulationResults/PlotData/results_MAC4_Multipath_1_64_CEErr_HBF_MC_2_bits_8_DBF_SNR_SE.txt};

          	\addplot[very thick, blue, dash dot, mark=x, mark repeat*=5, mark options={solid}] table[x = x, y=y]{./SimulationResults/PlotData/results_MAC4_Multipath_1_64_CEErr_HBF_MC_16_bits_1_DBF_SNR_SE.txt};
          	\addplot[very thick, BTORMIX16, dash dot, mark=x, mark repeat*=5, mark options={solid}] table[x = x, y=y]{./SimulationResults/PlotData/results_MAC4_Multipath_1_64_CEErr_HBF_MC_16_bits_2_DBF_SNR_SE.txt};
          	\addplot[very thick, BTORMIX26, dash dot, mark=x, mark repeat*=5, mark options={solid}] table[x = x, y=y]{./SimulationResults/PlotData/results_MAC4_Multipath_1_64_CEErr_HBF_MC_16_bits_3_DBF_SNR_SE.txt};
          	\addplot[very thick, BTORMIX36, dash dot, mark=x, mark repeat*=5, mark options={solid}] table[x = x, y=y]{./SimulationResults/PlotData/results_MAC4_Multipath_1_64_CEErr_HBF_MC_16_bits_4_DBF_SNR_SE.txt};
          	\addplot[very thick, BTORMIX46, dash dot, mark=x, mark repeat*=5, mark options={solid}] table[x = x, y=y]{./SimulationResults/PlotData/results_MAC4_Multipath_1_64_CEErr_HBF_MC_16_bits_5_DBF_SNR_SE.txt};
          	\addplot[very thick, BTORMIX56, dash dot, mark=x, mark repeat*=5, mark options={solid}] table[x = x, y=y]{./SimulationResults/PlotData/results_MAC4_Multipath_1_64_CEErr_HBF_MC_16_bits_6_DBF_SNR_SE.txt};
          	\addplot[very thick, BTORMIX66, dash dot, mark=x, mark repeat*=5, mark options={solid}] table[x = x, y=y]{./SimulationResults/PlotData/results_MAC4_Multipath_1_64_CEErr_HBF_MC_16_bits_7_DBF_SNR_SE.txt};
          	\addplot[very thick, red, dash dot, mark=x, mark repeat*=5, mark options={solid}] table[x = x, y=y]{./SimulationResults/PlotData/results_MAC4_Multipath_1_64_CEErr_HBF_MC_16_bits_8_DBF_SNR_SE.txt};
          	
          	\node[anchor=west]  at (axis cs:-5,26){increasing resolution};
   		\node (source) at (axis cs:20,12.5){};  		
   		\node (destination) at (axis cs:20,25){};
   		\draw[->, very thick](source)--(destination);
          	
     \end{axis}
\end{tikzpicture}
\vspace*{-0.5cm}
\caption{HBF average achievable rate for $M_R = 64$, $U = 4$, $M_{\text{RFE}} \in \{ 4, 32\}$ and ADC resolution $b \in \{1, \cdots, 8\}$.}
\label{fig:SNRSEHBF}
\end{figure}
\begin{figure}
\centering
\begin{tikzpicture}
    \begin{axis}[width=0.95*8.8cm, height=0.95*8.8cm, ylabel={avg. achievable rate [bps/Hz]}, grid,  legend cell align=left, ymin=0, ymax=40,
                  		xlabel={SNR [dB]}, xmin =-30, xmax=30, legend pos=north west, legend cell align=left, legend columns=1]
                  		
             	\addlegendimage{very thick, blue, mark=square, mark options={solid}, dashed}
 			\addlegendentry{DBF mixed $M_h = 32$}   
 			\addlegendimage{very thick, blue, dashed, mark=x, mark options={solid}}
 			\addlegendentry{DBF mixed $M_h = 4$}   	
 			
          	\addplot[very thick, blue, mark=square, mark repeat*=5, mark options={solid}, dashed] table[x = x, y=y]{./SimulationResults/PlotData/results_MAC4_Multipath_1_64_CEErr_32_32_5bit__DBF_bits_1_DBF_SNR_SE.txt};
          	\addplot[very thick, BTORMIX12, mark=square, mark repeat*=5, mark options={solid}, dashed] table[x = x, y=y]{./SimulationResults/PlotData/results_MAC4_Multipath_1_64_CEErr_32_32_5bit__DBF_bits_2_DBF_SNR_SE.txt};
          	\addplot[very thick, BTORMIX22, mark=square, mark repeat*=5, mark options={solid}, dashed] table[x = x, y=y]{./SimulationResults/PlotData/results_MAC4_Multipath_1_64_CEErr_32_32_5bit__DBF_bits_3_DBF_SNR_SE.txt};
          	\addplot[very thick, red, mark=square, mark repeat*=5, mark options={solid}, dashed] table[x = x, y=y]{./SimulationResults/PlotData/results_MAC4_Multipath_1_64_CEErr_32_32_5bit__DBF_bits_4_DBF_SNR_SE.txt};
          	
          	\addplot[very thick, blue, dashed, mark=x, mark repeat*=5, mark options={solid}] table[x = x, y=y]{./SimulationResults/PlotData/results_MAC4_Multipath_1_64_CEErr_60_4_5bit__DBF_bits_1_DBF_SNR_SE.txt};
          	\addplot[very thick, BTORMIX12, dashed, mark=x, mark repeat*=5, mark options={solid}] table[x = x, y=y]{./SimulationResults/PlotData/results_MAC4_Multipath_1_64_CEErr_60_4_5bit__DBF_bits_2_DBF_SNR_SE.txt};
          	\addplot[very thick, BTORMIX22, dashed, mark=x, mark repeat*=5, mark options={solid}] table[x = x, y=y]{./SimulationResults/PlotData/results_MAC4_Multipath_1_64_CEErr_60_4_5bit__DBF_bits_3_DBF_SNR_SE.txt};
          	\addplot[very thick, red, dashed, mark=x, mark repeat*=5, mark options={solid}] table[x = x, y=y]{./SimulationResults/PlotData/results_MAC4_Multipath_1_64_CEErr_60_4_5bit__DBF_bits_4_DBF_SNR_SE.txt};
          	
          	\node[anchor=west]  at (axis cs:-5,10){increasing resolution};
   		\node (source) at (axis cs:20,11){};  		
   		\node (destination) at (axis cs:20,26.5){};
   		\draw[->, very thick](source)--(destination);
     \end{axis}
\end{tikzpicture}
\vspace*{-0.5cm}
\caption{DBF mixed average achievable rate for $M_R = 64$, $U = 4$, $M_{h} \in \{4, 32\}$ and ADC resolution $b \in \{1, \cdots, 8\}$.}
\label{fig:SNRSEDBFMixed}
\end{figure}
\fi
\ifCLASSOPTIONdraftcls
\begin{figure}
\centering
\begin{tikzpicture}
\begin{axis}[width=0.95*8.8cm, height=0.95*8.8cm, xlabel={avg. achievable sum-rate [bps/Hz]}, grid, xmax=4.5, ymax=5.5, xmin=0.5, ymin=0.9, legend pos=outer north east, legend cell align=left,
                  		ylabel={energy efficiency [bps/J]}]
		\addplot[very thick, red, mark=square, mark options={solid}] table[x = y, y=x]{./SimulationResults/PlotData/results_MAC4_Multipath_1_64_CEErr__DBF_bits__SNR_-150_DBF_SE_EE.txt};
		\addlegendentry{DBF}   
		\addplot[very thick, dash dot, blue, mark=x, mark options={solid}] table[x = y, y=x]{./SimulationResults/PlotData/results_MAC4_Multipath_1_64_CEErr_HBF_MC_16_bits__SNR_-150_HBF.txt};
		\addlegendentry{HBF $M_{\text{RFE}} = 4$}   
		\addplot[very thick, dash dot, BTOGMIX12, mark=o, mark options={solid}] table[x = y, y=x]{./SimulationResults/PlotData/results_MAC4_Multipath_1_64_CEErr__HBF_MC_8_bits__SNR_-150_HBF.txt};
		\addlegendentry{HBF $M_{\text{RFE}} = 8$}    
		\addplot[very thick, dash dot, BTOGMIX22, mark=triangle, mark options={solid}] table[x = y, y=x]{./SimulationResults/PlotData/results_MAC4_Multipath_1_64_CEErr_HBF_MC_4_bits__SNR_-150_HBF.txt};
		\addlegendentry{HBF $M_{\text{RFE}} = 16$}   
		\addplot[very thick, dash dot, green, mark=square, mark options={solid}] table[x = y, y=x]{./SimulationResults/PlotData/results_MAC4_Multipath_1_64_CEErr_HBF_MC_2_bits__SNR_-150_HBF.txt};
		\addlegendentry{HBF $M_{\text{RFE}} = 32$}    
		
		\node[anchor=west]  at (axis cs:1.5,1.1){increasing resolution};
   		\node (source) at (axis cs:1.5,1.25){};  		
   		\node (destination) at (axis cs:2.5,2.5){};
   		\draw[->, very thick](source)--(destination);
		
     \end{axis}
\end{tikzpicture}
\vspace*{-0.5cm}
\caption{Spectral and energy efficiency of digital and hybrid beamforming with $M_R = 64$, $U = 4$, $M_{\text{RFE}} \in \{4, 8, 16, 32\}$ and ADC resolution $b \in \{1, \cdots, 8\}$ at SNR -15 dB.}
\label{fig:SEEE-15}
\end{figure}
\begin{figure}
\centering
\begin{tikzpicture}
\begin{axis}[width=0.95*8.8cm, height=0.95*8.8cm, xlabel={avg. achievable sum-rate [bps/Hz]}, grid, xmax=20, ymax=22.5, xmin=2.5, ymin=4, legend pos=outer north east, legend cell align=left,
ylabel={energy efficiency [bps/J]}]
		\addplot[very thick, red, mark=square, mark options={solid}] table[x = y, y=x]{./SimulationResults/PlotData/results_MAC4_Multipath_1_64_CEErr__DBF_bits__SNR_0_DBF_SE_EE.txt};
		\addlegendentry{DBF}    
		\addplot[very thick, dash dot, blue, mark=x, mark options={solid}] table[x = y, y=x]{./SimulationResults/PlotData/results_MAC4_Multipath_1_64_CEErr_HBF_MC_16_bits__SNR_0_HBF.txt};
		\addlegendentry{HBF $M_{\text{RFE}} = 4$}    
		\addplot[very thick, dash dot, BTOGMIX12, mark=o, mark options={solid}] table[x = y, y=x]{./SimulationResults/PlotData/results_MAC4_Multipath_1_64_CEErr__HBF_MC_8_bits__SNR_0_HBF.txt};
		\addlegendentry{HBF $M_{\text{RFE}} = 8$}    
		\addplot[very thick, dash dot, BTOGMIX22, mark=triangle, mark options={solid}] table[x = y, y=x]{./SimulationResults/PlotData/results_MAC4_Multipath_1_64_CEErr_HBF_MC_4_bits__SNR_0_HBF.txt};
		\addlegendentry{HBF $M_{\text{RFE}} = 16$}    
		\addplot[very thick, dash dot, green, mark=square, mark options={solid}] table[x = y, y=x]{./SimulationResults/PlotData/results_MAC4_Multipath_1_64_CEErr_HBF_MC_2_bits__SNR_0_HBF.txt};
		\addlegendentry{HBF $M_{\text{RFE}} = 32$}    
		
		\node[anchor=west]  at (axis cs:7.5,6){increasing resolution};
   		\node (source) at (axis cs:7.5,6.5){};  		
   		\node (destination) at (axis cs:13,13){};
   		\draw[->, very thick](source)--(destination);
     \end{axis}
\end{tikzpicture}
\vspace*{-0.5cm}
\caption{Spectral and energy efficiency of digital and hybrid beamforming with $M_R = 64$, $U = 4$, $M_{\text{RFE}} \in \{4, 8, 16, 32\}$ and ADC resolution $b \in \{1, \cdots, 8\}$ at SNR 0 dB.}
\label{fig:SEEE0}
\end{figure}
\begin{figure}
\centering
\begin{tikzpicture}
\begin{axis}[width=0.95*8.8cm, height=0.95*8.8cm, xlabel={avg. achievable sum-rate [bps/Hz]}, grid, xmax=35, ymax=35, xmin=2.5,ymin=4.5, legend pos=outer north east, legend cell align=left,
ylabel={energy efficiency [bps/J]}]
		\addplot[very thick, red, mark=square, mark options={solid}] table[x = y, y=x]{./SimulationResults/PlotData/results_MAC4_Multipath_1_64_CEErr__DBF_bits__SNR_150_DBF_SE_EE.txt};
		\addlegendentry{DBF}   
		\addplot[very thick, dash dot, blue, mark=x, mark options={solid}] table[x = y, y=x]{./SimulationResults/PlotData/results_MAC4_Multipath_1_64_CEErr_HBF_MC_16_bits__SNR_150_HBF.txt};
		\addlegendentry{HBF $M_{\text{RFE}} = 4$}   
		\addplot[very thick, dash dot, BTOGMIX12, mark=o, mark options={solid}] table[x = y, y=x]{./SimulationResults/PlotData/results_MAC4_Multipath_1_64_CEErr__HBF_MC_8_bits__SNR_150_HBF.txt};
		\addlegendentry{HBF $M_{\text{RFE}} = 8$}   
		\addplot[very thick, dash dot, BTOGMIX22, mark=triangle, mark options={solid}] table[x = y, y=x]{./SimulationResults/PlotData/results_MAC4_Multipath_1_64_CEErr_HBF_MC_4_bits__SNR_150_HBF.txt};
		\addlegendentry{HBF $M_{\text{RFE}} = 16$}   
		\addplot[very thick, dash dot, green, mark=square, mark options={solid}] table[x = y, y=x]{./SimulationResults/PlotData/results_MAC4_Multipath_1_64_CEErr_HBF_MC_2_bits__SNR_150_HBF.txt};
		\addlegendentry{HBF $M_{\text{RFE}} = 32$}   
		
		\node[anchor=west]  at (axis cs:7.5,6){increasing resolution};
   		\node (source) at (axis cs:7.5,7){};  		
   		\node (destination) at (axis cs:17.5,17.5){};
   		\draw[->, very thick](source)--(destination);
     \end{axis}
\end{tikzpicture}
\vspace*{-0.5cm}
\caption{Spectral and energy efficiency of digital and hybrid beamforming with $M_R = 64$, $U = 4$, $M_{\text{RFE}} \in \{4, 8, 16, 32\}$ and ADC resolution $b \in \{1, \cdots, 8\}$ at SNR 15 dB.}
\label{fig:SEEE15}
\end{figure}
\begin{figure}
\centering
\begin{tikzpicture}
\begin{axis}[width=0.95*8.8cm, height=0.95*8.8cm, xlabel={avg. achievable sum-rate [bps/Hz]}, grid, xmin=15, xmax=35, ymin=15, ymax=35, ylabel={energy efficiency [bps/J]}, legend pos=outer north east, legend cell align=left]
		\addplot[very thick, red, mark=square, mark options={solid}] table[x = y, y=x]{./SimulationResults/PlotData/results_MAC4_Multipath_1_64_CEErr__DBF_bits__SNR_150_DBF_SE_EE.txt};
		\addlegendentry{DBF}  
		\addplot[very thick, dashed, blue, mark=x, mark options={solid}] table[x = y, y=x]{./SimulationResults/PlotData/results_MAC4_Multipath_1_64_CEErr_60_4_5bit__DBF_bits__SNR_150_DBF_SE_EE.txt};
		\addlegendentry{DBF mixed $M_h = 4$}  
		\addplot[very thick, dashed, BTOGMIX12, mark=o, mark options={solid}] table[x = y, y=x]{./SimulationResults/PlotData/results_MAC4_Multipath_1_64_CEErr_56_8_5bit__DBF_bits__SNR_150_DBF_SE_EE.txt};	
		\addlegendentry{DBF mixed $M_h = 8$}  	
		\addplot[very thick, dashed, BTOGMIX22, mark=triangle, mark options={solid}] table[x = y, y=x]{./SimulationResults/PlotData/results_MAC4_Multipath_1_64_CEErr_48_16_5bit__DBF_bits__SNR_150_DBF_SE_EE.txt};
		\addlegendentry{DBF mixed $M_h = 16$}  
		\addplot[very thick, dashed, green, mark=square, mark options={solid}] table[x = y, y=x]{./SimulationResults/PlotData/results_MAC4_Multipath_1_64_CEErr_32_32_5bit__DBF_bits__SNR_150_DBF_SE_EE.txt};
		\addlegendentry{DBF mixed $M_h = 32$}  

		\node[anchor=west]  at (axis cs:17,32){increasing resolution};
   		\node (source) at (axis cs:20,26){};  		
   		\node (destination) at (axis cs:27,32){};
   		\draw[->, very thick](source)--(destination);
     \end{axis}
\end{tikzpicture}
\vspace*{-0.5cm}
\caption{Spectral and energy efficiency of digital beamforming with/without mixed ADC configuration with $M_R = 64$, $U = 4$, $M_{\text{h}} \in \{4, 8, 16, 32\}$ and ADC resolution $b \in \{1, \cdots, 8\}$, $b_l \in \{1, \cdots, 4\}$ and $b_h = 5$ at SNR 15 dB.}
\label{fig:SEEEmixed15}
\end{figure}
\else
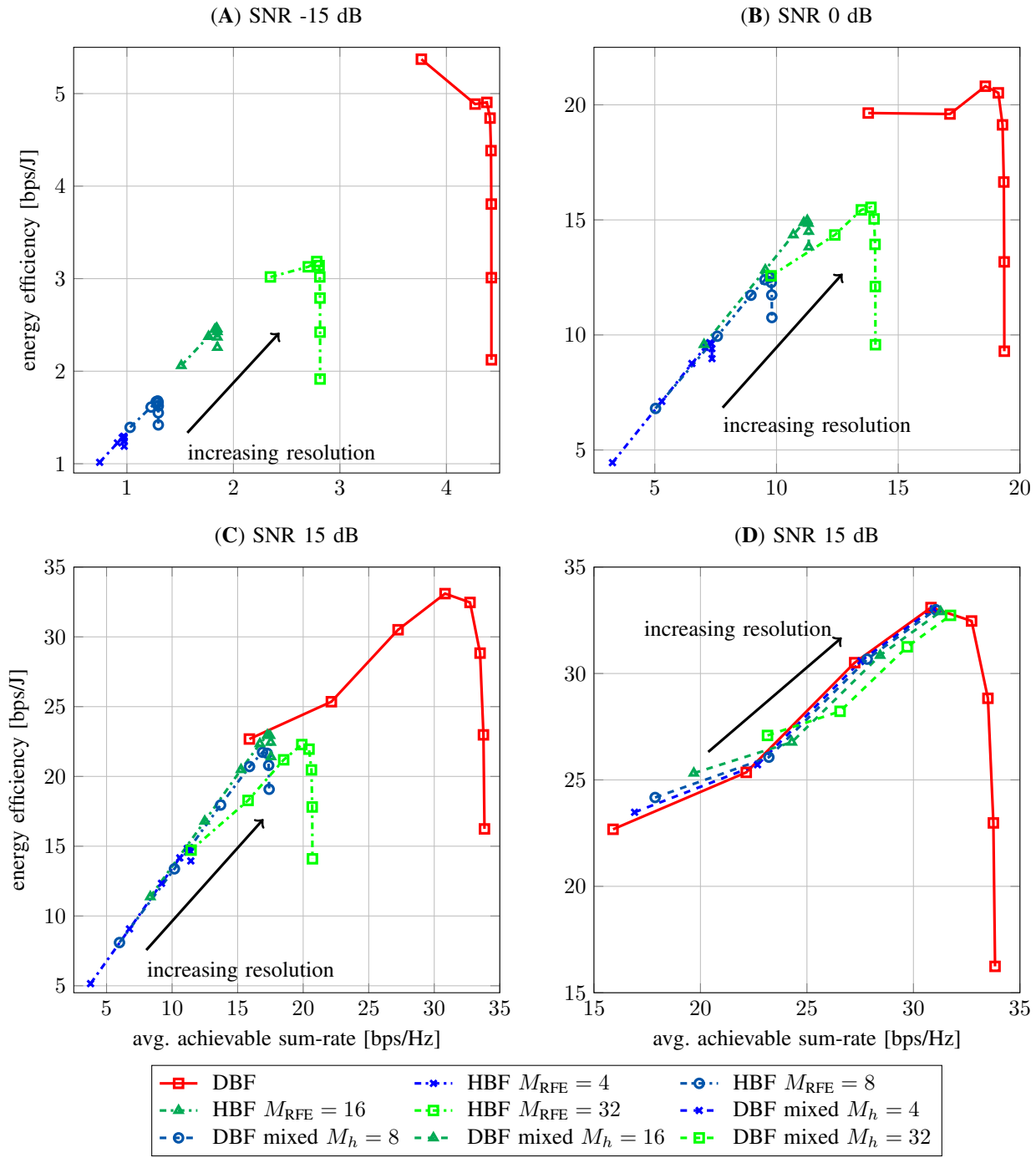
\begin{figure*}[!t]
\begin{center}
\normalsize
\begin{tikzpicture}
\begin{groupplot}[group style={group size= 2 by 2, horizontal sep=1.5cm, vertical sep=1.5cm}, width=18.1cm, height=18.1cm,
]
    \nextgroupplot[title={(\textbf{A}) SNR -15 dB}, width=0.95*8.8cm, height=0.95*8.8cm, grid, xmax=4.5, ymax=5.5, xmin=0.5, ymin=0.9,
                  		ylabel={energy efficiency [bps/J]}]
		\addplot[very thick, red, mark=square, mark options={solid}] table[x = y, y=x]{./SimulationResults/PlotData/results_MAC4_Multipath_1_64_CEErr__DBF_bits__SNR_-150_DBF_SE_EE.txt};
		\addplot[very thick, dash dot, blue, mark=x, mark options={solid}] table[x = y, y=x]{./SimulationResults/PlotData/results_MAC4_Multipath_1_64_CEErr_HBF_MC_16_bits__SNR_-150_HBF.txt};
		\addplot[very thick, dash dot, BTOGMIX12, mark=o, mark options={solid}] table[x = y, y=x]{./SimulationResults/PlotData/results_MAC4_Multipath_1_64_CEErr__HBF_MC_8_bits__SNR_-150_HBF.txt};
		\addplot[very thick, dash dot, BTOGMIX22, mark=triangle, mark options={solid}] table[x = y, y=x]{./SimulationResults/PlotData/results_MAC4_Multipath_1_64_CEErr_HBF_MC_4_bits__SNR_-150_HBF.txt};
		\addplot[very thick, dash dot, green, mark=square, mark options={solid}] table[x = y, y=x]{./SimulationResults/PlotData/results_MAC4_Multipath_1_64_CEErr_HBF_MC_2_bits__SNR_-150_HBF.txt};  
		
		\coordinate (c1) at (rel axis cs:0,1);
		\node[anchor=west]  at (axis cs:1.5,1.1){increasing resolution};
   		\node (source) at (axis cs:1.5,1.25){};  		
   		\node (destination) at (axis cs:2.5,2.5){};
   		\draw[->, very thick](source)--(destination);
   		
   		 \nextgroupplot[title={(\textbf{B}) SNR 0 dB}, width=0.95*8.8cm, height=0.95*8.8cm, grid, xmax=20, ymax=22.5, xmin=2.5, ymin=4]
		\addplot[very thick, red, mark=square, mark options={solid}] table[x = y, y=x]{./SimulationResults/PlotData/results_MAC4_Multipath_1_64_CEErr__DBF_bits__SNR_0_DBF_SE_EE.txt};
		\addplot[very thick, dash dot, blue, mark=x, mark options={solid}] table[x = y, y=x]{./SimulationResults/PlotData/results_MAC4_Multipath_1_64_CEErr_HBF_MC_16_bits__SNR_0_HBF.txt}; 
		\addplot[very thick, dash dot, BTOGMIX12, mark=o, mark options={solid}] table[x = y, y=x]{./SimulationResults/PlotData/results_MAC4_Multipath_1_64_CEErr__HBF_MC_8_bits__SNR_0_HBF.txt};
		\addplot[very thick, dash dot, BTOGMIX22, mark=triangle, mark options={solid}] table[x = y, y=x]{./SimulationResults/PlotData/results_MAC4_Multipath_1_64_CEErr_HBF_MC_4_bits__SNR_0_HBF.txt};  
		\addplot[very thick, dash dot, green, mark=square, mark options={solid}] table[x = y, y=x]{./SimulationResults/PlotData/results_MAC4_Multipath_1_64_CEErr_HBF_MC_2_bits__SNR_0_HBF.txt};
		
		\coordinate (c2) at (rel axis cs:1,1);
		\node[anchor=west]  at (axis cs:7.5,6){increasing resolution};
   		\node (source) at (axis cs:7.5,6.5){};  		
   		\node (destination) at (axis cs:13,13){};
   		\draw[->, very thick](source)--(destination);
   		
   		\nextgroupplot[title={(\textbf{C}) SNR 15 dB}, width=0.95*8.8cm, height=0.95*8.8cm, xlabel={avg. achievable sum-rate [bps/Hz]}, grid, xmax=35, ymax=35, xmin=2.5,ymin=4.5, 
   		ylabel={energy efficiency [bps/J]}]
		\addplot[very thick, red, mark=square, mark options={solid}] table[x = y, y=x]{./SimulationResults/PlotData/results_MAC4_Multipath_1_64_CEErr__DBF_bits__SNR_150_DBF_SE_EE.txt};
		\addplot[very thick, dash dot, blue, mark=x, mark options={solid}] table[x = y, y=x]{./SimulationResults/PlotData/results_MAC4_Multipath_1_64_CEErr_HBF_MC_16_bits__SNR_150_HBF.txt};
		\addplot[very thick, dash dot, BTOGMIX12, mark=o, mark options={solid}] table[x = y, y=x]{./SimulationResults/PlotData/results_MAC4_Multipath_1_64_CEErr__HBF_MC_8_bits__SNR_150_HBF.txt};
		\addplot[very thick, dash dot, BTOGMIX22, mark=triangle, mark options={solid}] table[x = y, y=x]{./SimulationResults/PlotData/results_MAC4_Multipath_1_64_CEErr_HBF_MC_4_bits__SNR_150_HBF.txt};
		\addplot[very thick, dash dot, green, mark=square, mark options={solid}] table[x = y, y=x]{./SimulationResults/PlotData/results_MAC4_Multipath_1_64_CEErr_HBF_MC_2_bits__SNR_150_HBF.txt}; 
		
		\node[anchor=west]  at (axis cs:7.5,6){increasing resolution};
   		\node (source) at (axis cs:7.5,7){};  		
   		\node (destination) at (axis cs:17.5,17.5){};
   		\draw[->, very thick](source)--(destination);
   		
   		\nextgroupplot[title={(\textbf{D}) SNR 15 dB}, width=0.95*8.8cm, height=0.95*8.8cm, xlabel={avg. achievable sum-rate [bps/Hz]}, grid, xmin=15, xmax=35, ymin=15, ymax=35, legend cell align=left,
   		legend to name=grouplegend2, legend style={legend columns=3,fill=none,draw=black,anchor=center,align=left, column sep=0.15cm}]
   		
   		\addlegendimage{very thick, red, mark=square, mark options={solid}}
		\addlegendentry{DBF}  
		
		\addlegendimage{very thick, dash dot, blue, mark=x, mark options={solid}}
		\addlegendentry{HBF $M_{\text{RFE}} = 4$} 
		\addlegendimage{very thick, dash dot, BTOGMIX12, mark=o, mark options={solid}}  
		\addlegendentry{HBF $M_{\text{RFE}} = 8$}
		\addlegendimage{very thick, dash dot, BTOGMIX22, mark=triangle, mark options={solid}}   
		\addlegendentry{HBF $M_{\text{RFE}} = 16$}   
		\addlegendimage{very thick, dash dot, green, mark=square, mark options={solid}}
		\addlegendentry{HBF $M_{\text{RFE}} = 32$}    
		
		\addlegendimage{very thick, dashed, blue, mark=x, mark options={solid}}
		\addlegendentry{DBF mixed $M_h = 4$} 
		\addlegendimage{very thick, dashed, BTOGMIX12, mark=o, mark options={solid}} 
		\addlegendentry{DBF mixed $M_h = 8$} 
		\addlegendimage{very thick, dashed, BTOGMIX22, mark=triangle, mark options={solid}} 	
		\addlegendentry{DBF mixed $M_h = 16$}  
		\addlegendimage{very thick, dashed, green, mark=square, mark options={solid}}
		\addlegendentry{DBF mixed $M_h = 32$} 
   		
		\addplot[very thick, red, mark=square, mark options={solid}] table[x = y, y=x]{./SimulationResults/PlotData/results_MAC4_Multipath_1_64_CEErr__DBF_bits__SNR_150_DBF_SE_EE.txt};
		\addplot[very thick, dashed, blue, mark=x, mark options={solid}] table[x = y, y=x]{./SimulationResults/PlotData/results_MAC4_Multipath_1_64_CEErr_60_4_5bit__DBF_bits__SNR_150_DBF_SE_EE.txt};
		\addplot[very thick, dashed, BTOGMIX12, mark=o, mark options={solid}] table[x = y, y=x]{./SimulationResults/PlotData/results_MAC4_Multipath_1_64_CEErr_56_8_5bit__DBF_bits__SNR_150_DBF_SE_EE.txt};	
		\addplot[very thick, dashed, BTOGMIX22, mark=triangle, mark options={solid}] table[x = y, y=x]{./SimulationResults/PlotData/results_MAC4_Multipath_1_64_CEErr_48_16_5bit__DBF_bits__SNR_150_DBF_SE_EE.txt};
		\addplot[very thick, dashed, green, mark=square, mark options={solid}] table[x = y, y=x]{./SimulationResults/PlotData/results_MAC4_Multipath_1_64_CEErr_32_32_5bit__DBF_bits__SNR_150_DBF_SE_EE.txt};

		\node[anchor=west]  at (axis cs:17,32){increasing resolution};
   		\node (source) at (axis cs:20,26){};  		
   		\node (destination) at (axis cs:27,32){};
   		\draw[->, very thick](source)--(destination);
\end{groupplot}
    	\coordinate (c3) at ($(c1)!.5!(c2)$);
    	\node[below] at (c3 |- current bounding box.south)
	{\ref{grouplegend2}};
\end{tikzpicture}
\vspace*{-0.3cm}
\caption{Spectral and energy efficiency of digital beamforming with/without mixed ADC configuration and hybrid beamforming with $M_R = 64$, $U = 4$, $M_{\text{RFE}} \in \{4, 8, 16, 32\}$, $M_{\text{h}} \in \{4, 8, 16, 32\}$ and ADC resolution $b \in \{1, \cdots, 8\}$, $b_l \in \{1, \cdots, 4\}$ and $b_h = 5$ at SNR $\in \{-15 \text{dB}, 0 \text{dB}, 15 \text{dB}\}$.}
\label{fig:SEEECombined}
\hrulefill
\end{center}
\vspace*{-0.1cm}
\end{figure*}
\fi

\section{Conclusion}
The evaluations in this paper showed that low resolution \ac*{ADC} digital beamforming systems are more energy efficient and achieves a higher rate than hybrid
 beamforming systems for multiuser scenario. 
 The reason is that the sub-arrays of hybrid beamforming must focus on a single user.
Evaluations with mixed \ac*{ADC} configurations showed that such systems can achieve different achievable rate and energy efficiency values around the ones achieve by a uniform \ac*{ADC} configuration. 
 
Future extensions should consider the following points. 
For the hybrid beamforming case, the evaluation only shows the result if the beams are already aligned. As shown in \cite{CellSearchDirectionalmmW},
beam alignment can require a large overhead. In addition considering what degree of power disparity among the users is possible for different \ac*{ADC} resolutions also provides a interesting scenario to evaluate.

\appendix  
\label{appendixAQNM}
In \cite{PaperDSP2015} we showed how to calculate the output correlation of a quantized system from the input correlation of a Gaussian signal. However in \cite{KILIANJSACEE} we assumed, that the same uniform quantizer is used
for the signal at each antenna. Here we generalize this result to include non-uniform quantization. We also combine this result with the results in \cite{AQNMAMINE} to include the effects of the quantization into the rate calculation.
The formula for calculating the quantizer output correlation from the input correlation for a general quantizer and two zero mean Gaussian random variables $a$ and $c$ can be written as
\begin{equation}
\label{eq_cross}
\rho_o = \sum\limits^{N_a - 1}_{l=1} \sum\limits^{N_c - 1}_{j=1} a^r_l c^r_j \int\limits_{0}^{\rho'_i} f_{ac}(a_l^s,c_j^s, \rho_i)~d\rho_i,
\end{equation}
with the joint probability density function $f_{ac}$ defined as
\begin{equation}
\begin{gathered}
f_{ac}(a, c, \rho_i) = \\
\frac{1}{2\pi\sigma_a\sigma_c\sqrt{1-\rho_i^2}}\exp\left(-\frac{1}{2(1-\rho_i^2)}\left[\frac{a^2}{\sigma_a^2} + \frac{c^2}{\sigma_c^2} - \frac{2\rho_i ac}{\sigma_a\sigma_c}\right]\right).
\end{gathered}
\end{equation}
The quantizers used for $a$ and $c$ have $N_a$ and $N_c$ quantization levels. The symbols $a^r_l$, $c^r_l$, $a^s_l$ and $c^s_j$ value of the quantization bins and the positions of the steps. 
It is important to mention that we assume that the representatives of the quantization bins and the position of the steps are adapted to the input power. In a practical system this is done by a \ac*{AGC} loop. 
Since we need to perform this transformation for every antenna pair, calculating the integral for every point is a large overhead.
Therefore, we generated a non-uniform grid of input correlation $\rho'_i$ in the range from 0 to $1$ and calculate the corresponding output correlation $\rho_o$ via numeric integration. 
The points in the grid are chosen in way that the change of $\rho_o$ between adjacent points in the grid do not exceed a threshold. Afterwards if we need to calculate the output correlation for a specific input correlation, 
we use the pre-calculated points and interpolate with cubic splines between them. This approach provides sufficient accuracy with reduced complexity.

With this technique we can calculate the correlation matrix after the quantization $\boldsymbol{R}_{\boldsymbol{r}\boldsymbol{r}}$ from the correlation matrix before the quantization $\boldsymbol{R}_{\boldsymbol{y}\boldsymbol{y}}$.
This procedure consists of the calculation of the diagonal elements of the matrix as 
\begin{equation}
	\left[\boldsymbol{R}_{\boldsymbol{r}\boldsymbol{r}}\right]_{i,i} = (1 - \sigma^2_{qi})\left[\boldsymbol{R}_{\boldsymbol{y}\boldsymbol{y}}\right]_{i,i},
\end{equation}
where $\sigma^2_{qi}$ is the variance of the distortion introduced by the quantization.
For each off diagonal element we use the formula in Equation \eqref{eq_cross} for all combinations of real and imaginary parts to calculate the resulting element in $\boldsymbol{R}_{\boldsymbol{r}\boldsymbol{r}}$.
We combine this procedure to form the operator $\text{T}(\cdot)$
\begin{equation}
	\boldsymbol{R}_{\boldsymbol{r}\boldsymbol{r}} = \text{T}\left(\boldsymbol{R}_{\boldsymbol{y}\boldsymbol{y}}, Q^1(\cdot), \cdots, Q^{M_{\text{RFE}}}\right).
\end{equation}
As shown in \cite{AQNMAMINE}, the matrix $\boldsymbol{F}$ for the Bussgang decomposition is a diagonal matrix. For the case of a different quantizer at each antenna the $i$th diagonal element is defined as
\begin{equation}
	\left[\boldsymbol{F}\right]_{i,i} = (1 - \sigma_{qi}).
\end{equation}
We can combine this operation with an operator \text{TF} only dependent on the quantization step functions $Q^i(\cdot)$
\begin{equation}
	\boldsymbol{F} = \text{TF}(Q^1(\cdot), \cdots, Q^{M_{\text{RFE}}}).
\end{equation}

%


\section*{Acknowledgment}
This work has been performed in the framework of the Horizon 2020 project ONE5G (ICT-760809) receiving funds from the European Union. The authors would like to acknowledge the contributions of their colleagues in the project, although the views expressed in this contribution are those of the authors and do not necessarily represent the project.



\bibliographystyle{IEEEtran}
\bibliography{./literature/IEEEabrv,./literature/bibKilian_locale}

%
%
%
%
%




\end{document}